\documentclass[icol,sn-nature,super]{sn-jnl}
\usepackage{graphicx}
\usepackage{multirow}%
\usepackage{amsmath,amssymb,amsfonts}%
\usepackage{amsthm}%
\usepackage{mathrsfs}%
\usepackage[title]{appendix}%
\usepackage{xcolor}%
\usepackage{textcomp}%
\usepackage{manyfoot}%
\usepackage{booktabs}%
\usepackage{algorithm}%
\usepackage{algorithmicx}%
\usepackage{algpseudocode}%
\usepackage{listings}%
\usepackage{anyfontsize}
\usepackage[version=4]{mhchem}
\usepackage{comment}
\usepackage{lineno}
\usepackage{setspace}
\makeatletter
\long\def\@makecaption#1#2{%
  \vskip\abovecaptionskip
  \normalsize 
  \textbf{#1.} #2\par
  \vskip\belowcaptionskip}
\makeatother
\newcommand{\cofirstdag}{$^{\dagger}$}
\begin{document}
\title{\textbf{Sliding multiferroicity in hexagonal stacked \ce{CrI3}}}

\author[1]{\fnm{Carter} \sur{Fox}\cofirstdag}
\author[2]{\fnm{Jose D.} \sur{Mella}\cofirstdag}
\author[3]{\fnm{Jack} \sur{Rollins}}
\author[3]{\fnm{Yangchen} \sur{He}}
\author[4]{\fnm{Yulu} \sur{Mao}}
\author[4]{\fnm{Haotian} \sur{Jiang}}
\author[3]{\fnm{Alaina} \sur{Drew}}
\author[4]{\fnm{Hongrui} \sur{Ma}}
\author[5]{\fnm{Takashi} \sur{Taniguchi}}
\author[6]{\fnm{Kenji} \sur{Watanabe}}
\author[1,3,4]{\fnm{Ying} \sur{Wang}}
\author[3]{\fnm{Daniel} \sur{Rhodes}}
\author*[2]{\fnm{Salvador} \sur{Barraza-Lopez}}\email{sbarraza@uark.edu}
\author*[1,3,4]{\fnm{Jun} \sur{Xiao}}\email{jun.xiao@wisc.edu}

\affil[1]{\orgdiv{Department of Physics}, \orgname{University of Wisconsin-Madison}, \orgaddress{\city{Madison},  \state{WI}, \country{USA}}}

\affil[2]{\orgdiv{Department of Physics}, \orgname{University of Arkansas}, \orgaddress{\city{Fayetteville}, \state{AR}, \country{USA}}}

\affil[3]{\orgdiv{Department of Materials Science and Engineering}, \orgname{University of Wisconsin-Madison}, \orgaddress{\city{Madison}, \state{WI}, \country{USA}}}

\affil[4]{\orgdiv{Department of Electrical and Computer Engineering}, \orgname{University of Wisconsin-Madison}, \orgaddress{\city{Madison}, \state{WI}, \country{USA}}}

\affil[5]{\orgdiv{Research Center for Materials Nanoarchitectonics}, \orgname{National Institute for Materials Science}, \orgaddress{\city{Tsukuba}, \state{Tsukuba}, \country{Japan}}}

\affil[6]{\orgdiv{Research Center for Electronic and Optical Materials}, \orgname{National Institute for Materials Science}, \orgaddress{\city{Tsukuba}, \state{Tsukuba}, \country{Japan}}}

\affil[$^{\dagger}$]{These authors contributed equally to this work}

\maketitle
\section*{Abstract}
\textbf{Developing new multiferroics at the two-dimensional (2D) limit with energy-efficient magnetoelectric coupling can inform the interplay physics of novel orders and advance on-chip high-performance computing applications. Here we apply stacking order engineering to create a new type of 2D multiferroics, namely sliding multiferroics, based on polar hexagonal stacked (H-stacked) few-layer \ce{CrI3}. This new stacking order removes structural inversion symmetry and gives rise to room temperature sliding ferroelectricity, as confirmed by Raman spectroscopy, second harmonic generation spectroscopy and electrical transport measurements. Building upon the gate-dependent reflective magnetic circular dichroism, first-principles calculations, and modeling, sliding ferroelectricity is shown to interplay with an emergent interfacial ferromagnetism via interlayer spin-polarized charge transfer. This coupling mechanism results in non-volatile magnetic switching by as low as 0.4~V across the H-stacked \ce{CrI3}. Our demonstration introduces polar stacking order engineering of 2D magnets as a general approach to create non-volatile 2D multiferroics with efficient magnetoelectric coupling, paving the way for low-power electronics and spintronics at the atomically thin limit.}

\newpage
\section*{Main}
Multiferroic materials with coexisting ferroic orders, such as ferroelectricity and ferromagnetism, offer an exciting platform for exploring the coupling between non-volatile quantum orderings and developing energy-efficient multifunctional electronics\cite{fiebig_evolution_2016,zhang_magnetoelectric_2025,spaldin_advances_2019,shen_new_2024-multiferroics}. 
In particular, multiferroics with nontrivial magnetoelectric (ME) coupling can enable non-volatile electric field control of magnetism, which can be significantly more efficient than magnetic field control\cite{spaldin_advances_2019}, and may have far reaching implications for innovative low-power information processing technologies such as multi-state computing\cite{wang_electric-field-driven_2020} and spintronics\cite{ahn_2d_2020}. 
Traditionally, complex oxides such as \ce{BiFeO3} along with engineered heterostructures have been extensively investigated and shown multiferroic behavior\cite{shen_new_2024-multiferroics,heron_deterministic_2014,spaldin_advances_2019,manipatruni_scalable_2019}, but face key challenges: the scarcity of single-phase multiferroics arising from the inherent \textit{d}-orbital incompatibility between ferroelectricity and magnetism, and heterostructure design restricted by lattice-mismatch constraints. 
Furthermore, the substantial depolarization fields and interfacial roughness set a lower limit on device thickness, often requiring switching voltages of several volts\cite{Ramesh2024}. 

Recent advances in van der Waals (vdW) materials have revealed  robust ferroelectrics and ferromagnets with atomically clean interfaces down to the ultimate two-dimensional limit\cite{chang2016,huang_layer-dependent_2017,xiao-dipolelock} ($\lesssim1$~nm). Moreover, their interlayer vdW bonding nature further allows on-demand homo- and hetero-structure fabrication without lattice mismatch limitations. 
Despite their great potential, only a few vdW multiferroics have been experimentally realized yet with limited functionality. For example, \ce{NiI2} shows evidence of multiferroicity at the monolayer level\cite{song_evidence_nii2-2022}; however, the magnetic ground state is a complex spin spiral that is highly sensitive and has no net magnetization \cite{li_realistic_nii2-2023,NiI2-shear-stab}, which is not compatible with the established on-chip electrical reading schemes\cite{song_giant_2018,spintronics-jia,ahn_2d_2020}. 
Multiferroic heterostructures have also been engineered by proximity, i.e., by combining layers of a 2D ferroelectric with a 2D ferromagnet, such as \ce{CuCrP2S6} (CCPS) coupled with \ce{Fe3GeTe2} (FGT)\cite{liang_preprint}. There, switching the CCPS ferroelectric state requires only $\sim2.2$~V; however, the thickness remains in the tens of nanometers and the switching only changes the magnetic coercive field, rather than directly changing the instantaneous magnetization. 

In light of these challenges and reliant on vdW stacking order engineering\cite{fox_stacking_2024}, a new type of multiferroic has recently been proposed based on endowing 2D magnets with sliding ferroelectricity\cite{zhong_theoretical_2023,Bennet-magents-multiferroic,wu_sliding_2021,gen_theory_stack_ferro}. Sliding ferroelectricity is a novel behavior in vdW materials in which the stacking order breaks mirror ($m_z$) and inversion ($\mathcal{P}$) symmetries, allowing for a new electric dipole moment that can be switched by interlayer sliding\cite{xiao_berry_2020,yasuda_stacking-engineered_2021,wang_interfacial_2022,wu_sliding_2021,fei_ferroelectric_2018,fox_stacking_2024}. In this proposed strategy, inherently non-polar monolayers of 2D magnets (e.g., \ce{CrI3}, \ce{Fe3GeTe2}, \ce{MnBi2Te4}, etc.) are artificially brought into non-centrosymmetric stacking orders that can host sliding ferroelectricity\cite{gen_theory_stack_ferro,zhong_theoretical_2023,VS2magnetoelec}. This strategy has been successfully employed to impart sliding ferroelectricity to non-magnetic 2D materials from their non-polar monolayers, including transition metal dichalcogenides (TMDs)\cite{wang_interfacial_2022} and hexagonal boron nitride (hBN)\cite{yasuda_stacking-engineered_2021,vizner_stern_interfacial_2021}. 
Coupling sliding ferroelectricity with the rich spin orders of 2D magnets via polar stacking order engineering remains experimentally unexplored. If successful, it would enable a general approach to create a novel group of ultrathin {\em sliding multiferroics} for energy-efficient nanoscale electronics. Besides, the vdW sliding barrier (few meV per formula unit) is one to two orders of magnitude smaller than the covalent/ionic bonding barrier for conventional ferroelectric switching\cite{wu_sliding_2021}, which may enable low operation voltage towards 100~mV for non-volatile electric-field control of magnetism, a long-sought goal for the multiferroic community\cite{spaldin_advances_2019}. 

In this article, we demonstrate a sliding multiferroic material by stacking few-layer \ce{CrI3} flakes with a 180$^\circ$ twist (hexagonal-stacked), forming a polar stacking order that simultaneously exhibits interfacial ferromagnetism and room temperature sliding ferroelectricity. Combining various optical and electrical characterizations with first-principles calculations and models, we reveal the nontrivial ME coupling through interlayer spin-polarized charge transfer tied to sliding ferroelectric switching. The new coupling mechanism, not observed in natural \ce{CrI3}, enables non-volatile magnetic switching even at electrical operating voltages below 0.4~V across the hexagonal-stacked \ce{CrI3}.

\section*{Realization of polar hexagonal stacked \ce{CrI3}}

\begin{figure*}[htb]
    \centering
    \includegraphics[width=\linewidth]{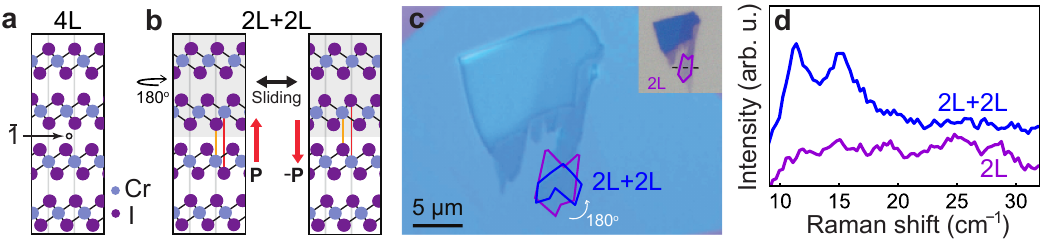} 
    \caption{
    \textbf{Polar stacking order engineering of \ce{CrI3}}. \textbf{a,} Side view of the four-layer (4L) natural monoclinic \ce{CrI3} with an inversion center ($\bar{1}$) located between the middle layers. Blue balls represent \ce{Cr} atoms while purple balls represent \ce{I} atoms. \textbf{b,} Side view of the crystal structure of a 2L+2L hexagonal stacked \ce{CrI3} (\ce{H-CrI3}). In this stacking, the top two layers have been twisted by $180^\circ$, breaking $\mathcal{P}$ and allowing for a vertical electric dipole $P_z$ pointing upward (left) or downward (right panel). The two energy-degenerate configurations are switchable by an interlayer sliding displacement between the top and bottom bilayers. \textbf{c,} Optical image of a typical 2L+2L \ce{H-CrI3} sample, where the two bilayer flakes are outlined in purple and blue. The inset shows the original flake for ``tear-and-stack,'' where the bilayer is outlined in purple and the dashed line marks the approximate tearing location. \textbf{d,} Ultralow-frequency Raman spectrum of  2L+2L \ce{H-CrI3} (blue) compared to a natural monoclinic 2L sample (purple) at 295 K. The two emergent interlayer phonon modes with strong Raman activity are attributed to the symmetry reduction in the polar \ce{H-CrI3}.}\label{fig:structure}
\end{figure*}

Figure \ref{fig:structure}a shows the structure of four \ce{CrI3} layers in their natural monoclinic stacking \cite{sun_giant_2019}. The spins within each layer are ferromagnetically (FM) coupled with an out-of-plane easy axis, while adjacent layers couple antiferromagnetically (AFM)\cite{huang_layer-dependent_2017,exchange_interlayer,exchange_intralayer}. 
The magnetic layer group (LG) is $c2/m'11$ ($c2'/m'11$) for even (odd) number of layers. Above the N\'eel temperature ($T_N\sim45$~K)\cite{huang_layer-dependent_2017}, the LG is the centrosymmetric $c2/m$. The natural \ce{CrI3} is nonpolar, which prevents the presence of any ferroelectricity.

To break this limitation, we create the polar hexagonal stacking order by introducing a 180$^\circ$ interlayer twist between two few-layer flakes. Specifically, we fabricate \ce{CrI3} samples from mechanically exfoliated flakes by the ``tear-and-stack'' method\cite{cao_superlattice-induced_2016} (See Methods for more details). The number of layers is determined by optical contrast ({\bf Supplementary Fig.~1}).  Such a stacking order, referred to as hexagonal stacked \ce{CrI3} (\ce{H-CrI3}), has been predicted to be a polar structure allowing for finite out-of-plane electrical polarization, switchable via interlayer sliding \cite{poudel_creating_2023,sun_theoretical_2023,zhong_theoretical_2023}. For instance, Figure \ref{fig:structure}b illustrates a 2L+2L \ce{H-CrI3} sample. The polar stacking induces an electric dipole, whose vertical component is represented by arrows in Figure~\ref{fig:structure}b and reverses direction upon sliding\cite{poudel_creating_2023}. There are a range of theoretical predictions of the vertical dipole in the 1L+1L case, up to $\sim1.5$~pC/m\cite{sun_theoretical_2023,poudel_creating_2023,zhong_theoretical_2023}. In {\bf Supplementary Fig.~2} we show the structures for 5L {\em vs.} polar 2L+3L and 6L {\em vs.} polar 3L+3L. Besides, the adjacent layers at the stacked interface are FM coupled \cite{poudel_creating_2023,sun_theoretical_2023,zhong_theoretical_2023} (the 1L+1L stack belongs to the non-centrosymmetric magnetic LG $cm'11$ (non-magnetic LG $cm11$) below (above) $T_N$). 

Figure \ref{fig:structure}c is an optical image of one typical 2L+2L \ce{H-CrI3} sample, with the two  bilayer components outlined to highlight the H-stacked area. The inset shows the original flake with the bilayer (2L) outlined in purple. Building upon the successful fabrication of \ce{H-CrI3}, we first conduct ultralow-frequency Raman spectroscopy to examine the corresponding Raman activity of interlayer phonon modes (Fig. \ref{fig:structure}d), which was proven to be sensitive to the stacking order symmetry\cite{Cheon2021_transition_mote2,Zhang2015grapheneraman,xiao_berry_2020}. Specifically, we compare the low-frequency Raman spectrum of a 2L+2L \ce{H-CrI3} sample against that of 2L natural monoclinic \ce{CrI3} at room temperature (295~K). Although there is a theoretical prediction of Raman active shear and breathing modes\cite{rodriguez-vega_phonon-mediated_2020} in 2Ls, such modes have not been experimentally seen in a previous report\cite{cenker_direct_2021} nor in our measurements (purple curve in figure \ref{fig:structure}d), which suggest their negligible Raman activity in the natural centrosymmetric monoclinic stacking order. On the other hand, two peaks emerge at 11.4~cm$^{-1}$ and 15.2~cm$^{-1}$ in the 2L+2L \ce{H-CrI3} sample, identified by fitting the data to a double Lorentzian function. These peaks likely correspond to an interlayer shear mode and an interlayer breathing mode, and are found to persist down to 2~K ({\bf Supplementary Fig.~3a}). As they are prevalent in the paramagnetic phase (above $T_N$), their strong Raman activity is a consequence of the polar hexagonal stack breaking $\mathcal{P}$ rather than magnetic ordering. Two vibrational modes at $\sim10.5$ cm$^{-1}$ and $\sim20$ cm$^{-1}$ in the spectrum of 1L+1L \ce{CrI3} were predicted \cite{poudel_creating_2023}; the discrepancy between the frequencies observed in 2L+2L samples may be explained by the different number of layers, as shear and breathing modes are thickness-sensitive\cite{Lorchat-raman-rese2res2}.
The emergent low-frequency modes highlight the symmetry reduction by the polar hexagonal stacking. 

\section*{Observation of  sliding ferroelectricity in hexagonal stacked \ce{CrI3}}

\begin{figure*}[htb]
    \centering
    \includegraphics[width=\linewidth]{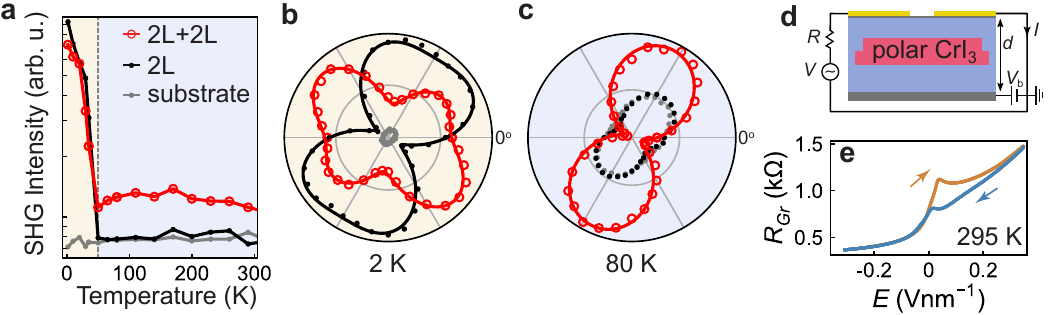}
    \caption{\textbf{Stacking order asymmetry and sliding ferroelectricity in hexagonal-stacked \ce{CrI3}}.  \textbf{a,} Temperature dependence of the second harmonic generation (SHG) intensity (in log scale) excited by linear polarized light for 2L+2L \ce{H-CrI3} and 2L natural monoclinic \ce{CrI3} samples. Both samples are paramagnetic above $T_N\sim $45 K (light blue shaded temperature regime) and  a negligible SHG is found in the natural monoclinic \ce{CrI3} flake due to its centrosymmetric crystal structure. In contrast, the \ce{H-CrI3} sample shows a much stronger SHG that persists up to room temperature, resulting from its polar stacking order. Below $T_N\sim $45~K (light yellow shaded regime), the long-range spin order in both samples breaks $\mathcal{P}$ and dominates the SHG contribution.  \textbf{b},\textbf{c}, Polarization-resolved SHG intensity as a function of the incident polarization angle below  (b, at 2 K) and above (c, at 80 K) the magnetic transition temperature for the same 2L+2L \ce{H-CrI3} and 2L natural monoclinic \ce{CrI3} samples in (a). The distinct SHG patterns observed for \ce{H-CrI3} reflect a different space group arising from its polar stacking order (see {\bf Supplementary Note 1} for more details). Data for 2L+2L, 2L, and the substrate are in red, black and grey, respectively. Solid lines in subplots (b) and (c) are SHG fits from the appropriate (magnetic or nonmagnetic) point group.
    \textbf{d,} Schematic of a device in which the resistance of the top few-layer graphene senses the local electric field change induced by the out-of-plane dipole switching in \ce{H-CrI3}. 
    \textbf{e,} Resistance of the top few-layer graphene ($R_{Gr}$) as the vertical electric field ($E=V_b/d$) is swept up and down in a loop for a 2L+2L stacked device at 295~K, showing hysteretic behavior indicative of sliding ferroelectricity.}
    \label{fig:SHG}
\end{figure*}

To further examine the structural asymmetry and the emergence of a switchable out-of-plane electric dipole in \ce{H-CrI3}, we conducted optical second-harmonic generation (SHG) spectroscopy and electrical transport measurements. The former has been widely used as a sensitive probe of lattice asymmetry and space group in layered materials with various stacking orders\cite{xiao_berry_2020,xiao-dipolelock}. We use a reflection geometry at normal incidence with linearly polarized 1040 nm ultrafast pulses (see Methods). Figure \ref{fig:SHG}a shows the SHG intensity at a fixed incident polarization angle as a function of temperature for natural monoclinic 2L \ce{CrI3} (black) and 2L+2L \ce{H-CrI3} (red). For 2L \ce{CrI3}, a large SHG response is only observed below $T_N\sim45$~K, as expected for $\mathcal{P}$ breaking by the AFM interlayer order~\cite{sun_giant_2019,GUDELLI2020896}. It becomes paramagnetic above $T_N$ and develops a center of inversion, making the SHG response negligible. In contrast, the 2L+2L \ce{H-CrI3} has a nontrivial SHG response above $T_N$ even up to room temperature, as the polar hexagonal stacking remains non-centrosymmetric in the paramagnetic phase. 

Furthermore, the space groups of the different stacking orders are revealed by the polarization-resolved SHG intensity patterns. Figure \ref{fig:SHG}b shows the angular dependence of the SHG response {\em versus} the incident polarization angle at 2~K ($<T_N$), while Figure \ref{fig:SHG}c displays the signal at 80~K ($>T_N$). The response of the 2L flake at 2~K is described by a $\chi^{(2)}$ tensor obtained from magnetic point group $2/m'$  (solid line). In contrast, the 2L+2L SHG signal has the symmetry of magnetic point group $m'$\cite{SHG_magnetic}. At 80 K, the 2L+2L stack is described by the nonmagnetic point group $m$, which still lacks $\mathcal{P}$. The point groups and fit appear in {\bf Supplementary Note 1}. Similar behavior is observed in other \ce{H-CrI3} samples (3L+3L and 3L+2L); see {\bf Supplementary Fig.~4}. 

Having established the structural asymmetry and the space group of \ce{H-CrI3}, we now investigate the predicted out-of-plane electric dipole and associated ferroelectric switching via interlayer sliding. To implement this study, we employed the graphene sensor approach used in previous sliding ferroelectric reports \cite{wang_interfacial_2022,fei_ferroelectric_2018,xiao_berry_2020,yasuda_stacking-engineered_2021}.  We fabricated devices with 2L+2L \ce{H-CrI3} sandwiched between hBN dielectric layers with a bottom graphite gate as well as a few-layer graphene top gate with two Au electrical contacts separated by 1.5~$\mu$m (Fig. \ref{fig:SHG}d). The back gate voltage ($V_b$) determines the vertical electric field ($E=V_b/d$ where $d$ is the total thickness of the hBN and \ce{CrI3} layers). 
The resistance of the few-layer graphene is sensitive to the local electric field via capacitive doping (see Methods for details). Figure \ref{fig:SHG}e shows the resistance of the top few-layer graphene $R_{Gr}$ as the vertical electric field $E$ is swept in a loop at 295~K. We observe a hysteretic behavior of $R_{Gr}$ that is indicative of ferroelectric switching of the 2L+2L \ce{CrI3} sample. The hysteresis is asymmetric about $E=0$, which is commonly observed in other sliding ferroelectrics\cite{wang_interfacial_2022,fei_ferroelectric_2018}, and likely due to intrinsic doping of the graphene and/or asymmetric trapped charges induced by the fabrication process. We observe gradual hysteresis, which is typically observed in artificially stacked sliding ferroelectrics, and is interpreted as the gradual switching of small individual domains with different pinning strengths \cite{wang_interfacial_2022,yasuda_stacking-engineered_2021}. Such a hysteresis characteristic is repeatably observed in other \ce{H-CrI3} devices; see {\bf Supplementary Fig.~5}. As the measurement is at room temperature (far above $T_N$), the hysteresis cannot be a result of magnetic order. Additionally, gate dependent Raman measurements show no change in the high-frequency intralayer vibrational modes ({\bf Supplementary Fig.~3b}), indicating the transition does not result from intralayer structural changes. To quantify the out-of-plane electric dipole, we analyze the resistance peak shift in the observed hysteresis \cite{wang_interfacial_2022,yasuda_stacking-engineered_2021} (see {\bf Supplementary Note 2 and Supplementary Fig. 6}). The extracted vertical dipole of about 5 pC/m is consistent with the range of theoretical predictions for polar stacked \ce{CrI3}\cite{sun_theoretical_2023}, which again confirms that the origin of the observed resistance hysteresis is the electric-field driven sliding ferroelectricity switching of \ce{H-CrI3}.

\section*{Emergent interlayer ferromagnetic coupling at hexagonal stacked interfaces}\label{sec:magnetism}

\begin{figure*}[htb]
    \centering
    \includegraphics[width=\linewidth]{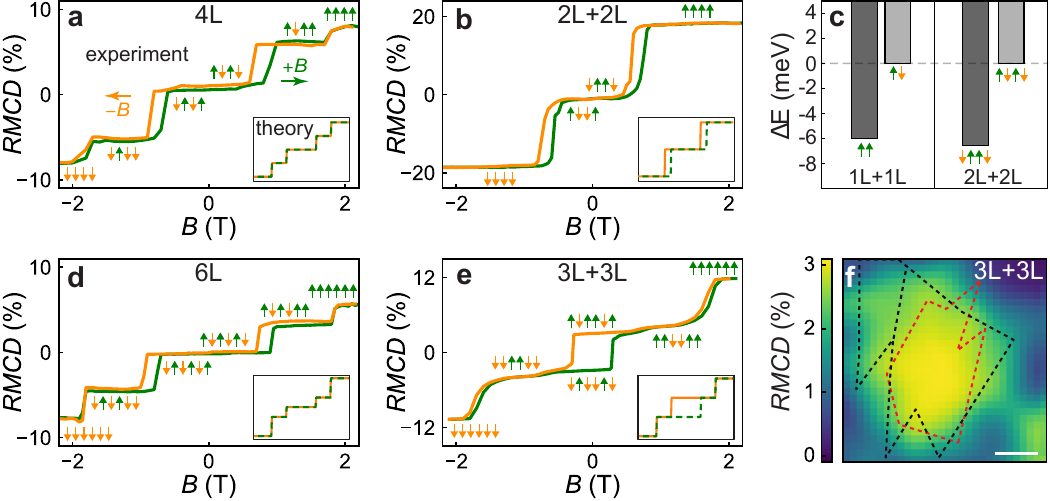}
    \caption{\textbf{Observation of  interfacial ferromagnetism in hexagonal-stacked \ce{CrI3}.} \textbf{a,b,} Reflective magnetic circular dichroism (RMCD) as a function of the applied out-of-plane $B$ for  (a) natural monoclinic 4L \ce{CrI3} and (b) 2L+2L \ce{H-CrI3}. \textbf{c,} Energy differences from first-pinciples calculations as the middle interface in 1L+1L (left) and 2L+2L \ce{H-CrI3} (right) are set to have AFM or FM order. $\Delta E$ denotes the change in energy relative to the AFM order in the middle interface. An FM order between the two layers in the middle of the polar stack is energetically preferred. \textbf{d,e,} RMCD as a function of out-of-plane $B$ for (d) 6L  and (e) 3L+3L  \ce{CrI3}. Insets in (a,b,d,e) show the simulated magnetization evolution based on an Ising model ({\bf Supplementary Note 3}) with emergent interfacial ferromagnetism, which agrees with the experimental observations. Orange and green arrows in subplots (a-e) denote the spin alignment in each layer. \textbf{f,} RMCD spatial mapping of a 3L+3L \ce{H-CrI3} sample, showing uniform interfacial ferromagnetism at $B=0$~T. The two trilayer flakes are outlined in red and black dashed lines.}
    \label{fig:magnetic}
\end{figure*}

With the presence of sliding ferroelectricity, we now study the coexisting magnetism in \ce{H-CrI3}. To this end, we perform reflective magnetic circular dichroism (RCMD) microscopy at 2~K on both natural and polar stacked samples. Figure \ref{fig:magnetic}a shows the RMCD {\em versus} the out-of-plane magnetic field $B$ for a natural 4L sample, which hosts interlayer AFM coupling between adjacent layers\cite{xie_evidence_2023,song_direct_2021}. From here on, $\uparrow$ and $\downarrow$ represent the magnetic moments of the out-of-plane spin configuration at each layer. The first transition occurs as an outer layer flips its magnetic moments at $B\sim\pm0.75$ T \cite{xie_evidence_2023,song_direct_2021} (i.e., from $\uparrow\downarrow\uparrow\downarrow$ to $\uparrow\downarrow\uparrow\uparrow$, or to $\downarrow\downarrow\uparrow\downarrow$): those flips require less energy than switching the magnetization of one of the inner two layers, which interface two other layers. A subsequent transition occurs at $B\sim\pm1.8$ T (i.e., from $\uparrow\downarrow\uparrow\uparrow$ to $\uparrow\uparrow\uparrow\uparrow$), when the $B$-field is large enough to flip the inner layer. In contrast, the RMCD curve for a 2L+2L \ce{H-CrI3} only displays one spin-flip transition at $B\sim\pm0.7$ T (Fig. \ref{fig:magnetic}b). This transition is from degenerate $\downarrow\uparrow\uparrow\downarrow$ or $\uparrow\downarrow\downarrow\uparrow$ magnetic ground states with zero magnetization onto fully aligned ferromagnetic states ($\uparrow\uparrow\uparrow\uparrow$ or $\downarrow\downarrow\downarrow\downarrow$, respectively). These degenerate ground states require interlayer FM coupling for \ce{H-CrI3} at the polar interface between the inner two layers while the outer layers retain an interlayer AFM coupling. Indeed, our DFT calculations of stacking energies for 1L+1L and 2L+2L in Figure \ref{fig:magnetic}c support FM interlayer alignment at polar stacked interfaces. Considering the $\downarrow\uparrow\uparrow\downarrow$ ground state, the outer two layers are expected to flip simultaneously onto $\uparrow\uparrow\uparrow\uparrow$ at the same critical $B-$field, which explains the observed single-step magnetic transition.

Such an interlayer FM coupling at polar hexagonal stacked interfaces is further corroborated by contrasting the magnetization curves of 6L and 3L+3L samples. Figure \ref{fig:magnetic}d shows the behavior of a 6L stack, displaying four magnetic transitions (at $B\sim \pm$0.75~T and $B\sim \pm$1.9~T) and an AFM ground state at $B=0$~T. In contrast, the 3L+3L stack in Figure~\ref{fig:magnetic}e exhibits a large remnant magnetization at $B=0$~T, consistent with either $\uparrow\downarrow\uparrow\uparrow\downarrow\uparrow$ or $\downarrow\uparrow\downarrow\downarrow\uparrow\downarrow$ ground states, in which the two layers at the middle of the polar stack continue to be FM coupled and the stack sustains a finite magnetization. Similar to Figure \ref{fig:magnetic}b, stacks with a FM polar stacked interface have less spin flip transitions (compare those in the 3L+3L stack against the 6L case, and see {\bf Supplementary Notes 3 and 4}). The observed FM interlayer alignment in \ce{H-CrI3} is also consistent with a recent demonstration of moir\'e  magnetism in H-stacked twisted \ce{CrI3} bilayers\cite{sun2025preprint}. 
The uniformity of the interfacial FM coupling was confirmed by a RMCD map over an entire 3L+3L sample at $B=0$~T (Fig. \ref{fig:magnetic}f). 

\newpage
\section*{Non-volatile magnetoelectric coupling in hexagonal-stacked \ce{CrI3} sliding multiferroics}

\begin{figure*}[htb]
    \centering
    \includegraphics[width=\linewidth]{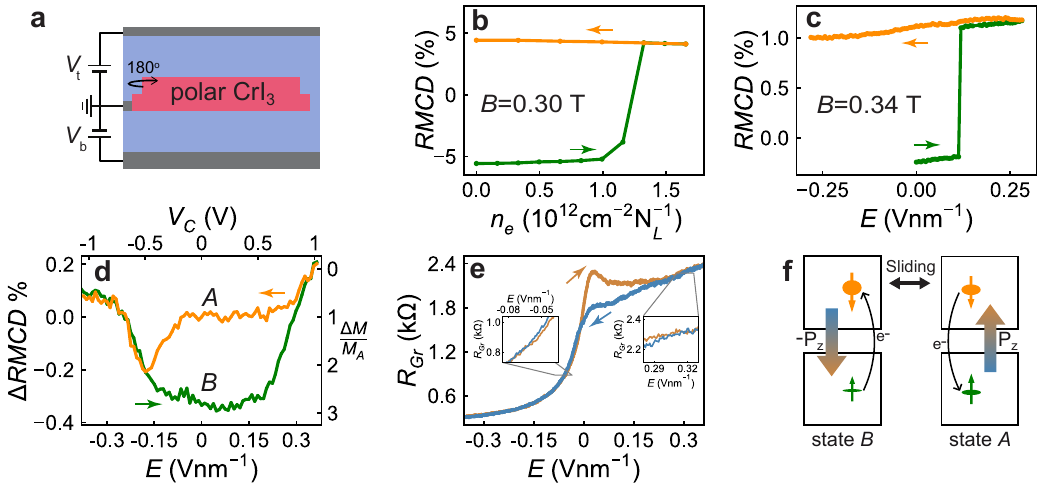}
    \caption{\textbf{Efficient non-volatile magnetoelectric coupling in  hexagonal-stacked \ce{CrI3} sliding multiferroics.} 
    \textbf{a,} Schematics of a dual-gate device based on \ce{H-CrI3}, which can independently control the external electron doping ($n_e$) and out-of-plane electric field ($E$) experienced by the \ce{H-CrI3}.  \textbf{b,}   $n_e$-assisted switching of the magnetic state of a 3L+3L \ce{H-CrI3} device at $B=0.30$~T, just below the coercive field. \textbf{c,} Out-of-plane electric field $E$-assisted switching of a 3L+2L \ce{H-CrI3} at $B=0.34$~T, just below the coercive field. 
    \textbf{d,} Pure out-of-plane electric field control of magnetism at $B=0$ T for a 2L+2L \ce{H-CrI3}. The non-volatile RMCD hysteresis suggests a nontrivial ME coupling in \ce{H-CrI3}. The left vertical axis shows the raw RMCD change relative to the RMCD of state $A$, whereas the right vertical axis shows the fractional change relative to state $A$.
    The top horizontal axis shows the estimated voltage across the \ce{CrI3}. 
    \textbf{e,} Top graphene resistance ($R_{Gr}$) as a function of $E$ in the same device as in subplot (d). The $R_{Gr}$ hysteresis is indicative of sliding ferroelectricity. Insets highlight where the forward and backward sweeps merge, indicating a hysteresis window similar to that of the magnetization in \textbf{d}. 
    Measurements in subplots (e) and (d) proceeded at $B=0$~T after ramping the magnetic field to $-2.2$~T and back to 0~T.
    \textbf{f,} Illustration of the proposed ME coupling mechanism in 2L+2L \ce{H-CrI3}. The non-volatile magnetism change is attributed to the interlayer spin-polarized charge transfer via the $E$-driven sliding ferroelectric transition in \ce{H-CrI3}.}
\label{fig:MEcoupling}
\end{figure*}

Building upon the coexistence of robust magnetism and sliding ferroelectricity in \ce{H-CrI3}, the next step is to investigate the ME coupling between these two ferroic orders. In particular, we study electrical control of magnetic switching in two regimes: (1) with the presence of finite $B$-fields close to the coercive field, and (2) without any applied $B$-field. To explore those regimes, we first fabricated dual-gated devices with graphite top and bottom gates, as well as a contact to the \ce{CrI3} (Fig. \ref{fig:MEcoupling}a). Electrostatic doping ($n_e$) and out-of-plane electric fields ($E$) can be independently controlled therein.

Figure~\ref{fig:MEcoupling}b illustrates $n_e$-driven switching of a 3L+3L device at $B=0.3$~T, i.e., just below the first critical $B$-field. This transition occurs at $n_e \sim 1.2 \times 10^{12}$~cm$^{-2}$~N$_L^{-1}$, where $N_L=6$ is the number of layers. The Ising model we developed ({\bf Supplementary Note 3}) suggests a transition in which the outer layers flip their magnetization (i.e., from $\downarrow\uparrow\downarrow\downarrow\uparrow\downarrow$ to $\uparrow\uparrow\downarrow\downarrow\uparrow\uparrow$). 
The equivalent energies within the Ising model of the two outer layers in the 3L+3L stack enables the $n_e$-driven magnetization change due to the two outer layers switching magnetic moments. 
The mechanism is suspected to be the interlayer exchange coupling between these natural stacking AFM interfaces decreasing with electron doping, lowering the critical $B$-field\cite{huang_electrical_2018,jiang_controlling_2018}.
The spin flip transition accessed in 3L+3L is also at considerably lower $B$-field than in the natural 6L case ({\bf Supplementary Note 4}).

Out-of-plane electric fields $E$ can also be used to switch the magnetization of 2L \ce{CrI3} \cite{jiang_electric-field_2018}, which is attributed to a linear ME coupling term altering the energy of the AFM ground state and thus the critical $B$-field for the spin-flip transition. In Figure \ref{fig:MEcoupling}c we show magnetic switching of a 3L+2L \ce{H-CrI3} device at $B=0.34$~T, just below the critical $B$-field. The transition from the $\downarrow \uparrow \downarrow \downarrow \uparrow$ state to the $\uparrow \uparrow \downarrow \downarrow \uparrow$ state involves one outer layer ({\bf Supplementary Fig.~7}) and requires $E\sim0.12$~V~nm$^{-1}$. In comparison, the $E$-field driven switching demonstrated in natural bilayers required $E\sim0.4$~V~nm$^{-1}$ at $B=0.4$~T\cite{jiang_electric-field_2018}. 
Additionally, since the thickness of a 3L+2L \ce{H-CrI3} is 3.5~nm, a voltage of the order of 1~V across the \ce{H-CrI3} is sufficient for magnetic switching. This is much smaller than the $\sim$~10~V needed to switch \ce{BiFeO3}\cite{Ramesh2024}, or the $\sim$~2.2~V used to change the magnetic coercive field in CCPS/FGT heterostructures electrostatically\cite{liang_preprint}.

Beyond $n_e$- and $E$-driven spin switching at finite $B$-fields, \ce{H-CrI3} can enable an efficient non-volatile electrical control of magnetism even without the need of external $B$-fields, which has not been demonstrated in natural monoclinically stacked \ce{CrI3}. As detailed below, such a non-volatile ME coupling is attributed to the unique interplay between their sliding ferroelectricity and an emergent magnetism mediated by interlayer spin-polarized charge transfer. To investigate their coupling, we revisit the device discussed in Figure \ref{fig:SHG}d-e, now at 2~K (i.e., in the magnetic order regime). We focus on the behavior at $B=0$~T after ramping $B$ down to $-2.2$~T and back to 0~T, which sets the polar stack in the $\uparrow\downarrow\downarrow\uparrow$ magnetic state (see `$+B$' trace in Fig.~\ref{fig:magnetic}b). We then measure the RMCD signal, which is proportional to the total out-of-plane magnetization, as a function of $E$ at $B=0$~T (Fig. \ref{fig:MEcoupling}d). We observe a bistable hysteresis and denote the two $E=0$ magnetization states as state $A$ and state $B$, where state $B$ has a larger negative magnetization than state $A$. In Figure \ref{fig:MEcoupling}d, the left vertical axis shows the change in RMCD relative to the RMCD of state $A$ ($M_A$), whereas the right vertical axis shows the fractional change with respect to $M_A$. 
The $E$-driven hysteresis of the RMCD signal at zero $B$-field displays a relative non-volatile magnetization change of about $\sim3\times$, meaning the magnetization of state $B$ is about three times that of state $A$. This observation is a clear hallmark of non-volatile ME coupling in \ce{H-CrI3}. The hysteresis is observed between $\sim-0.17$~Vnm$^{-1}$ and $\sim0.3$~Vnm$^{-1}$.  
To understand its microscopic origin, we also measured the top few-layer graphene resistance ($R_{Gr}$) as $E$ is varied in the same device (Fig. \ref{fig:MEcoupling}e). Similar to the room temperature case (Fig. \ref{fig:SHG}e), we observe a hysteretic behavior indicative of sliding ferroelectric switching of the 2L+2L \ce{H-CrI3} sample. Insets in Figure \ref{fig:MEcoupling}e highlight where the forward and backward sweeps merge at $\sim-0.06$~Vnm$^{-1}$ and $\sim0.3$~Vnm$^{-1}$, having a similar hysteresis window as the magnetization displayed in Figure~\ref{fig:MEcoupling}d. 
The synchronized hysteretic behaviors between the sliding ferroelectricity and magnetism indicate they may share the same out-of-plane electric field ($E$-driven) origin. 

Inspired by recent theoretical modeling of ME couplings in sliding multiferroics\cite{VS2magnetoelec,liu_tunable_2023,zhong_theoretical_2023,sun_theoretical_2023}, we propose that a spin redistribution arises from interlayer spin-polarized charge transfer via the $E$-driven sliding ferroelectric transition (Fig. \ref{fig:MEcoupling}d). Although charge transfer only between FM coupled layers would not alter the net magnetization of the 2L+2L stack, its magnitude may change upon charge redistribution between the top two and bottom two layers as the sliding ferroelectric switching occurs (see {\bf Supplementary Note 5} and {\bf Supplementary Figs.~8 and 9} for details). 
Figure \ref{fig:MEcoupling}f illustrates this concept by showing the two polarization states with the top and bottom bilayers represented by square blocks. The magnetization of each bilayer block is represented by green and orange arrows. Due to asymmetric charge distribution and/or the effect of the substrate, each bilayer may have a small inherent spin imbalance. For example, consider the case in Figure \ref{fig:MEcoupling}f where the top bilayer has a negative spin imbalance and the bottom bilayer has a smaller positive spin imbalance, therefore having a small negative net magnetization. 
Switching of the polarization from state $A$ to state $B$ (via sliding) involves spin polarized charge transfer from the top bilayer to the bottom bilayer, which lowers the magnitude of the net magnetization. 

Beyond the description of this novel ME coupling mechanism, we point out that the metrics of the non-volatile ME switching are appealing: the switching (coercive) $E$-fields for magnetism can be as low as $\sim-0.12$~V~nm$^{-1}$. Since the thickness of a 2L+2L stack is $t_c=2.8$~nm, this suggests that an intrinsic operating voltage smaller than 0.4~V could be achieved if the top and bottom gates are in direct contact with the stack. The estimated voltage across the 2L+2L stack ($V_C=E\cdot t_c$) is shown as the top horizontal axis of Figure \ref{fig:MEcoupling}d to highlight a potentially low operating voltage. Its $\sim0.4$~V magnitude is much smaller than the $\sim 10$ V needed to switch \ce{BiFeO3}\cite{Ramesh2024}. Even compared with existing vdW multiferroic platforms, such a low operating voltage is considerably smaller than the $\sim2.2$~V switching voltage for CCPS+FGT heterostructures, making ultrathin sliding multiferroics an ideal platform for next generation high-density low-energy electronics.

In summary, we employ stacking order engineering of 2D magnets to realize a new class of multiferroics, namely sliding multiferroics. Setting a polar (hexagonal) stacking of initially nonpolar magnetic 2D \ce{CrI3} bilayers and trilayers as a prototypical system, we demonstrate that $\mathcal{P}$ breaking gives rise to robust room-temperature sliding ferroelectricity, confirmed by multimodal electrical transport, Raman spectroscopy, and nonlinear optical probes. Moreover, we establish that sliding ferroelectricity couples directly to emergent interfacial ferromagnetism via interlayer spin-polarized charge transfer. This coupling enables non-volatile magnetic switching at a 0.4~V operation voltage, representing a long-sought benchmark for the multiferroics community. Our findings introduce polar stacking order engineering of 2D vdW magnets as a general approach to create 2D multiferroics with efficient ME coupling, paving the way for non-volatile low-power electronics and spintronics at the atomically thin limit.

\section*{Methods}\label{secMethods}

\textbf{\ce{CrI3} crystal synthesis.}
Bulk \ce{CrI3} crystals purchased from HQ Graphene were used to fabricate some of the samples; other crystals were grown with the  chemical vapor transport (CVT) method. With a total weight of 0.5 g, stoichiometric amounts of Cr powder (99.99$\%$, metals basis) and crystalline I (99.999$\%$, metals basis) were loaded into a quartz ampule (inner diameter 1.4 cm, length $\sim 13$ cm) and sealed under vacuum (10$^{-5}$ Torr). Subsequently, the sealed ampule was placed into a two-zone furnace and heated under a temperature gradient with the source material at the hot end. The hot (cold) end was ramped to 650 $^\circ$C (550 $^\circ$C) over a 24 h period and held at this temperature gradient for two weeks,  naturally cooling down by turning the furnace off. The process yielded shiny and flat \ce{CrI3} crystals at the cold end. \\ \\
\textbf{Sample fabrication.}
Polar stacked \ce{CrI3} samples were prepared by the layer-by-layer dry transfer method in a glovebox with $<0.01$ ppm \ce{O2} and \ce{H2O}. Thin flakes of \ce{CrI3}, hBN, and graphite were first mechanically exfoliated from bulk crystals onto 285 nm SiO$_2$/Si substrates. The number of layers of the \ce{CrI3} flakes was determined by optical contrast and confirmed by atomic force microscopy. The thickness of hBN dielectrics for dual-gated devices was determined by atomic force microscopy. 
Using a polymer stamp (PC or PPC), the ``tear-and-stack'' method was applied to stack few-layer \ce{CrI3} flakes with a $180^\circ$ twist, with hBN encapsulation on both sides taking place afterwards. For gated devices, thin graphite flakes were used for the top and bottom gates and for contact to the \ce{CrI3}. After completing the stack, Au contacts were made by standard e-beam lithography and metal deposition. \\ \\
\textbf{Optical measurements.}
Optical measurements were performed in a magneto-optical cryostat (Quantum Design, OptiCool) capable of operating between 1.7~K and 350~K. A femtosecond pulsed laser centered at 1040 nm with an 80 MHz repetition rate and $\sim$140 fs pulse duration (Coherent, Discovery NX) was focused onto the sample at normal incidence by a 50x long working distance objective for SHG measurements. The incident linear polarization angle was varied by rotating a half-wave plate in the incident beam path. The 520 nm SHG signal was split off by a dichroic mirror, isolated by band-pass filters, and detected by a photomultiplier tube (Hamamatsu, H11890).

For the RMCD measurements, a 633 nm HeNe laser was focused onto the sample at normal incidence, and the reflected light detected by a biased photodiode was passed to a lock-in amplifier (SSI-Instrument, OE1022d). The helicity of light was modulated between left and right circular polarization (LCP and RCP) at 100 kHz by a photoelastic modulator set to $\lambda/2$ retardation, followed by a quarter wave plate, and the intensity was modulated by a mechanical chopper at $\sim$6 kHz. The reflection signal was demodulated at the helicity modulation frequency to determine the difference between the reflected LCP and RCP intensity ($dR$), and at the chopper frequency to determine the average ($R$). The RMCD signal is defined as $dR/R$. 

Ultralow-frequency Raman spectroscopy measurements were conducted with a linearly polarized 633 nm HeNe laser focused onto the sample at normal incidence. The reflected light is passed through three band notch filters to filter out the Rayleigh peak before being directed to a spectrometer (Shamrock 500i, Andor) with an 1800 lines/mm grating. A half wave plate is rotated in the excitation path for polarization dependent measurements. \\ \\
\textbf{Dual-gated device control.}
We apply a simple parallel-plate capacitor model to control the vertical electric field ($E$) and external doping ($n_e$) in dual-gated devices with top and bottom hBN dielectric layers of thicknesses $d_t$ and $d_b$. 
For devices in which the \ce{CrI3} is grounded (Fig. \ref{fig:MEcoupling}a), the electric field and external doping are determined by $E=(-V_t/d_t + V_b/d_b)/2$ and $n_e=\epsilon_{BN}\epsilon_0(V_t/d_t+V_b/d_b)/e$, where $V_t$, $V_b$, $\epsilon_{BN}=4$, $\epsilon_0$ are the top gate bias, bottom gate bias, dielectric constant of hBN and vacuum permittivity, respectively. For the electric field sweeps, the top and bottom gate voltages are varied while keeping $-V_t/d_t=V_b/d_b$, whereas for external doping sweeps the top and bottom gate voltages are varied while keeping $V_t/d_t=V_b/d_b$. For devices without a contact to the \ce{CrI3} (Fig. \ref{fig:SHG}d), the top gate is grounded and the vertical electric field is determined by $E=V_b/d$, where $d$ is the total thickness of the hBN and \ce{CrI3} layers. The voltage across the \ce{CrI3} (top horizontal axis of Fig. \ref{fig:MEcoupling}d) is determined from the relative thickness of the \ce{CrI3} compared to the hBN layers. 
Source meters (Keithley, 2450/2400) were used to apply voltages to the top and bottom graphite gates in electric field and doping-dependent measurements. \\ \\
\textbf{Few-layer Graphene probe resistance measurement.}
Two Au contacts spaced by 1.5~$\mu$m were formed on the top few-layer graphene as source and drain electrodes to detect its resistance, which is sensitive to the local electric field in the top hBN layer. 
An AC voltage ($V$) is sourced by a lock-in amplifier (SSI-Instrument, OE1022d) and
a resistor ($R$) is placed in series with the top few-layer graphene. The voltage across the top graphene
($V_{Gr}$) is measured by the lock-in amplifier, demodulated at the source frequency.
The resistance of the top graphene is determined by $R_{Gr}$ = $V_{Gr}/I$, where the current is $I = (V - V_{Gr})/R$. For back gate sweeps, a source meter (Keithley, 2450) was used to apply DC voltages between the top and bottom graphite ($V_b$).\\ \\
\textbf{Computational methods.} DFT calculations were performed with the {\em Vienna Ab Initio Simulation Package} (VASP). \cite{vasp1} The effects of the vdW interaction were empirically included using the correction proposed by Grimme \textit{ et al.}\cite{vdW_DFT} We used a 15$\times$15$\times$1 $k$-point mesh, an energy cutoff of 500 eV, and an electronic energy convergence criteria of $10^{-6}$ eV. A force convergence of $10^{-4} ~\text{eV/\AA}$ was set for structural relaxations. Structural optimizations were carried without SOC, which was only turned on when structures were at local energy minima. To treat the strong on-site Coulomb interaction of Cr $d$ orbitals we used the DFT+U method introduced by Dudarev \textit{et al.},\cite{dudarev1998electron} with an effective value of $U$ equal to 3 eV ($J$ = 0) \cite{jiang2019stacking}. All calculations were performed with an out-of-plane lattice constant of $c\sim 60$ \AA~to reduce self-interaction among periodic copies. Convergence tests were performed to achieve an accuracy of the order of 1~meV in total energies. We computed the total energies of 1L+1L and 2L+2L stacks using DFT when the magnetic order in the polar-stacked middle interface was AFM and FM. We considered the magnetic order $\uparrow\downarrow$ and $\uparrow\uparrow$ for the 1L+1L case, and  $\uparrow\downarrow\uparrow\downarrow$ and $\uparrow\downarrow\downarrow\uparrow$ for the 2L+2L polar structure. The magnetic ground state is the magnetic configuration with the lowest energy.
\backmatter

\section*{Supplementary information} 
Supplementary Notes 1-5 contain analysis of SHG data, estimation of the vertical electric dipole in a 2L+2L stack at room temperature, an Ising model, a discussion of critical fields, and the properties of a 2L+2L stack under $E$. Supplementary Figs. 1-9 further support the claims made in the main text.

\section*{Acknowledgments}
C.F., J.R., A.D., and J.X.~acknowledge support by the U.S.~National Science Foundation CAREER program (Grant No.~DMR-2237761). J.D.M.~and S.B.L.~were funded by the NSF's Q-AMASE-i program (Award No.~DMR-1906383). Calculations were performed at the Arkansas High Performance Computing Center (NSF Award No.~OAC-2346752). We thank P.~Kumar for discussions. Y.H., Y.M., H. J., H. M., Y.W. and D.R. gratefully acknowledge partial support of this research by NSF through the University of Wisconsin Materials Research Science and Engineering Center (DMR-2309000). Y.W. acknowledges the additional support by the U.S.~National Science Foundation CAREER program (Grant No. ECCS-2239093). K.W. and T.T. acknowledge support from the JSPS KAKENHI (Grant Numbers 21H05233 and 23H02052), the CREST (JPMJCR24A5), JST and World Premier International Research Center Initiative (WPI), MEXT, Japan.

\section*{Author Contributions}

J.X.~conceived the research. J.X.~and C.F.~designed the experiments. S.B.L.~and J.X.~supervised the project. Y.H.~synthesized bulk high-quality \ce{CrI3} crystals under the guidance of D.R. T.T.~and K.W.~provided high-quality hBN bulk crystals. C.F. fabricated the devices with assistance from J.R., Y.M., H.J.~and A.D., under the guidance of Y.W.~and J.X.  Raman measurements were performed by C.F.~with assistance from H.M. In addition, C.F.~conducted SHG measurements, RMCD and electrical transport measurements, and analyzed the data with J.X. J.D.M.~performed DFT calculations and developed the models with feedback from C.F.~and S.B.L. All authors wrote the manuscript.

\section*{Competing interests}
The authors declare no competing interests. 

\section*{Data availability}
The data on this paper is available from the corresponding authors upon reasonable request.

\section*{Code availability}
The inputs used for the first-principles calculations and analytical modeling are available from the corresponding authors upon request.
\bigskip

\bibliography{newbib}

\clearpage
\renewcommand{\thefigure}{S\arabic{figure}}
\setcounter{figure}{0}

\begin{center}
    {\LARGE \bfseries Supplementary Information: Sliding multiferroicity in hexagonal stacked \ce{CrI3} \par}
    \vspace{1em}
    \normalsize
    Carter Fox$^{1,\dagger}$, Jose D. Mella$^{2,\dagger}$, Jack Rollins$^3$, Yangchen He$^3$, Yulu Mao$^4$, Haotian Jiang$^4$, Alaina Drew$^3$, Hongrui Ma$^4$,\\
    Takashi Taniguchi$^5$, Kenji Watanabe$^6$, Ying Wang$^{1,3,4}$, Daniel Rhodes$^3$,\\
    Salvador Barraza-Lopez$^{2,*}$, Jun Xiao$^{1,3,4,*}$
    
    \vspace{1em}
    $^1$Department of Physics, University of Wisconsin-Madison, Madison, WI, USA\\
    $^2$Department of Physics, University of Arkansas, Fayetteville, AR, USA\\
    $^3$Department of Materials Science and Engineering, University of Wisconsin-Madison, Madison, WI, USA\\
    $^4$Department of Electrical and Computer Engineering, University of Wisconsin-Madison, Madison, WI, USA\\
    $^5$Research Center for Materials Nanoarchitectonics, National Institute for Materials Science, Tsukuba, Japan\\
    $^6$Research Center for Electronic and Optical Materials, National Institute for Materials Science, Tsukuba, Japan\\
    $^\dagger$These authors contributed equally to this work.\\ 
    \vspace{1em}
    $^*$Corresponding author(s). E-mail(s): sbarraza@uark.edu, jun.xiao@wisc.edu
\end{center}

\section*{} 
\doublespacing
\noindent \textbf{Supplementary Note 1:} Analysis of SHG intensity pattern and stacking order space group \hfill \textbf{15} \\
\textbf{Supplementary Note 2:} Estimation of the out-of-plane electric dipole in 2L+2L \ce{H-CrI3} from experimental result \hfill \textbf{17} \\
\textbf{Supplementary Note 3:} Ising model simulation for emergent magnetism in \ce{H-CrI3} \hfill \textbf{19} \\
\textbf{Supplementary Note 4:} Critical field analysis and comparison between natural monoclinic and hexagonal-stacked \ce{CrI3} \hfill \textbf{19} \\
\textbf{Supplementary Note 5:} Spin-polarized charge transfer model in 2L+2L \ce{H-CrI3} under electric fields \hfill \textbf{20} \\
\textbf{Supplementary Figure S1:} Identifying the layer number of \ce{CrI3} flakes \hfill \textbf{24} \\
\textbf{Supplementary Figure S2:} Additional few-layer \ce{CrI3} crystal structures \hfill \textbf{24} \\
\textbf{Supplementary Figure S3:} Additional Raman spectrum at low temperature and with finite electric field at 2 K \hfill \textbf{25} \\
\textbf{Supplementary Figure S4:} SHG measurements of additional \ce{H-CrI3} samples at 80 K \hfill \textbf{25} \\
\textbf{Supplementary Figure S5:} Ferroelectric behavior in another \ce{H-CrI3} device \hfill \textbf{26} \\
\textbf{Supplementary Figure S6:} Estimating the out-of-plane electric dipole in 2L+2L \ce{H-CrI3} \hfill \textbf{26} \\
\textbf{Supplementary Figure S7:} Magnetic hysteresis modeling in 3L+2L \ce{H-CrI3} \hfill \textbf{27} \\
\textbf{Supplementary Figure S8:} Capacitor model for 2L+2L \ce{H-CrI3} \hfill \textbf{27} \\
\textbf{Supplementary Figure S9:} Sliding ferroelectric model simulation for 2L+2L \ce{H-CrI3} \hfill \textbf{28}

\newpage
\section*{1. Analysis of SHG intensity pattern and stacking order space group}
The second-order nonlinear optical response is described by the second-order polarization vector $\mathbf{P}^{(2)}(\phi)$ induced by a time-dependent applied electric field $\mathbf{E}(\phi)$ and given by $P_i(\phi)=\chi_{ijk}^{(2)}E_j(\phi)E_k(\phi)$, where $\chi_{ijk}^{(2)}=\chi_{ikj}^{(2)}$ is the second-order susceptibility tensor, subscripts $i,j,k \in \{x,y,z\}$ designate crystallographic axes, and $\phi$ is the angle between the incident polarization and the crystal's $x$-axis. For the sake of notation, we may redefine the angle as $\phi-\phi_0$ in the subsequent fits if required. Throughout this analysis, we assume normal incident light with linear in-plane polarization, such that its electric field is given by $\textbf{E}(\phi)=E_0(\cos(\phi),\sin(\phi),0)$. In the absence of a polarizer during detection, the second-harmonic generation (SHG) intensity is proportional to the in-plane $\textbf{P}^{(2)}(\phi)$ component, and is given by $I(\phi)\propto |P_x^{(2)}(\phi)|^2+|P_y^{(2)}(\phi)|^2$. We decompose $\chi^{(2)}_{ijk}$ into even and odd components under time reversal ($\mathcal{T}$):
\begin{equation*}
\chi^{(2)}_{ijk} = \chi^{(2)\text{ even}}_{ijk} + \chi^{(2)\text{ odd}}_{ijk}.
\end{equation*}
According to Neumann's principle, $\chi^{(2)}_{ijk}$ is constrained by the symmetries of the magnetic point group (PG). Magnetic symmetry operations include both unitary ($R$) and anti-unitary ($R'=\mathcal{T}R$) elements. Under $R$-type symmetry operations (with matrix elements $R_{ij}$), the even and odd parts of $\chi^{(2)}_{ijk}$ transform as:
\begin{eqnarray}\label{eq:R}
   \chi^{(2)\text{ even}}_{ijk} = \sum_{lmn} R_{il} R_{jm} R_{kn} \chi^{(2)\text{ even}}_{lmn},\\
   \chi^{(2)\text{ odd}}_{ijk} = \sum_{lmn} R_{il} R_{jm} R_{kn} \chi^{(2)\text{ odd}}_{lmn}.\nonumber
\end{eqnarray}

On the other hand, under $R'$-type operations (with matrix elements $R'_{ij}$), the even and odd parts transform as:
\begin{eqnarray}
\label{eq:Rp}
\chi^{(2)\text{ even}}_{ijk} = \sum_{lmn} R'_{il} R'_{jm} R'_{kn} \chi^{(2)\text{ even}}_{lmn},\\
\chi^{(2)\text{ odd}}_{ijk} = -\sum_{lmn} R'_{il} R'_{jm} R'_{kn} \chi^{(2)\text{ odd}}_{lmn}.\nonumber
\end{eqnarray}
$\chi_{ijk}^{(2)}$ is determined by Equation~\ref{eq:R} only when the material is described by a nonmagnetic PG. The substrate may contribute to the SHG, too. Let $\mathbf{P}_{\mathcal{G}_J}^{(2)}$ and $\mathbf{P}_{\mathrm{sub}}^{(2)}$ denote the second-order polarizations from a \ce{CrI3} stack and the substrate, respectively ($J=1,2,3$ represent the three different stacking configurations under consideration). The contribution of the substrate is treated as a constant that does not interfere with the \ce{CrI3} signal. Under this assumption, the SHG intensity is:
$I(\phi)\sim |\mathbf{P}_{\mathcal{G}_J}^{(2)}(\phi)|^2+|\mathbf{P}_{sub}^{(2)}|^2=I_{\mathcal{G}_J}(\phi)+I_{sub}$, with a constant $I_{sub}$.

$\mathcal{P}$ is broken by the interlayer magnetic configuration of CrI$_3$ bilayers: The 2L structure belongs to the magnetic PG
$\mathcal{G}_1=2/m'=\{\mathcal{I},\ C_{2y},\ \mathcal{T}\mathcal{P},\ \mathcal{T}m_y \}$, with $C_{2y}$ a twofold rotation about the $y$-axis, $m_y$ a mirror plane perpendicular to $y$, and $\mathcal{I}$ the identity. The $z$-axis corresponds to the out-of-plane direction. Equations \ref{eq:R} and \ref{eq:Rp} impose symmetry constraints under which the $\chi_{ijk}^{(2)\text{ even}}$ vanish. The $\chi_{ijk}^{(2)}$ tensor for PG $\mathcal{G}_1$ is (in Voigt notation):

\begin{equation*}
\begin{pmatrix}
0 & 0 & 0 & \chi^{(2)\text{ odd}}_{xyz} & 0 & \chi^{(2)\text{ odd}}_{xxy} \\
\chi^{(2)\text{ odd}}_{y x x} & \chi^{(2)\text{ odd}}_{yyy} & \chi^{(2)\text{ odd}}_{yzz} & 0 & \chi^{(2)\text{ odd}}_{y x z} & 0 \\
0 & 0 & 0 & \chi^{(2)\text{ odd}}_{z y z} & 0 & \chi^{(2)\text{ odd}}_{z x y} \\
\end{pmatrix}.
\end{equation*}
Its SHG intensity is expressed as:

\begin{equation*}
\begin{split}
I_{\mathcal{G}_1}(\phi) &\propto 4\cos^2(\phi)\sin^2(\phi)| \chi^{(2)\text{ odd}}_{xxy}|^2+| \chi^{(2)\text{ odd}}_{yxx} \cos^2\phi + \chi^{(2)\text{ odd}}_{yyy} \sin^2\phi |^2, \\ 
&=A\cos^4(\phi)
+ B\cos^2(\phi) \sin^2(\phi)
+ D\sin^4(\phi),
\end{split}
\end{equation*}

where $A=|\chi^{(2)\text{ odd}}_{yxx}|^2$, $B=2(2|\chi^{(2)\text{ odd}}_{xxy}|^2+Re(\chi^{(2)\text{ odd}}_{yxx}(\chi^{(2)\text{ odd}}_{yyy})^*))$, $D=|\chi^{(2)\text{ odd}}_{yyy}|^2$. We fitted the 2L curve shown in Figure 2b of the main text with $I_{sub}=0.3$, $\phi_0=0.54$ rads, $A= 0.066\pm 0.008$, $B=1.612 \pm 0.033$, and $D=0.648\pm0.008$ ($I_{sub}$, $A$, $B$, and $D$ are in arb.~u.). The 2L+2L stacks have a polar interface between the second and third layers. If $\mathbf{P}_1^{(2)}$, $\mathbf{P}_2^{(2)}$, and $\mathbf{P}_3^{(2)}$ denote nonlinear polarization at the lower, middle, and upper stacking interfaces, the total SHG signal is expressed as a coherent sum: $|\mathbf{P}_1^{(2)} + \mathbf{P}_2^{(2)} + \mathbf{P}_3^{(2)}|^2$. $\mathbf{P}_3^{(2)}$ is the inversion counterpart of $\mathbf{P}_1^{(2)}$. $\chi^{(2)}_{ijk}$ transforms as $\chi^{(2)}_{ijk} \rightarrow -\chi^{(2)}_{ijk}$ under a unitary transformation $\mathcal{P}$, which implies $\mathbf{P}_3^{(2)} = -\mathbf{P}_1^{(2)}$. As a result, the SHG intensity simplifies to $|\mathbf{P}_1^{(2)} + \mathbf{P}_2^{(2)} - \mathbf{P}_1^{(2)}|^2 \sim |\mathbf{P}_2^{(2)}|^2$, where $\mathbf{P}_2^{(2)}$ represents the contribution of a 1L+1L structure. This analysis was applied to both magnetic and nonmagnetic configurations. The interlayer magnetic order is FM for 1L+1L stacks with magnetic PG $m'=\mathcal{G}_2=\{\mathcal{I},\ \mathcal{T}m_y \}$, which allows both even and odd components. The $\chi_{ijk}^{(2)}$ tensor for PG $\mathcal{G}_2$ is (in Voigt notation):
\begin{equation*}
\begin{pmatrix}
\chi^{(2)\text{ even}}_{xxx} & \chi^{(2)\text{ even}}_{xyy} & \chi^{(2)\text{ even}}_{xzz} & \chi^{(2)\text{ odd}}_{xyz} & \chi^{(2)\text{ even}}_{xxz} & \chi^{(2)\text{ odd}}_{xxy} \\
\chi^{(2)\text{ odd}}_{y x x} & \chi^{(2)\text{ odd}}_{yyy} & \chi^{(2)\text{ odd}}_{yzz} & \chi^{(2)\text{ even}}_{yyz} & \chi^{(2)\text{ odd}}_{y x z} & \chi^{(2)\text{ even}}_{yxy} \\
\chi^{(2)\text{ even}}_{zxx} & \chi^{(2)\text{ even}}_{zyy} & \chi^{(2)\text{ even}}_{zzz} & \chi^{(2)\text{ odd}}_{z y z} & \chi^{(2)\text{ even}}_{zxz} & \chi^{(2)\text{ odd}}_{z x y} \\
\end{pmatrix}.
\end{equation*}

All $\chi^{(2)}_{ijk}$ are allowed, so ``even'' and ``odd'' superscripts are omitted for simplicity. The SHG intensity is:
\begin{equation*}
\begin{split}
I_{\mathcal{G}_2}(\phi) \propto\ 
&\left| \cos^2(\phi)\,\chi_{xxx}^{(2)} + \sin^2(\phi)\,\chi_{xyy}^{(2)} + 2\cos(\phi)\sin(\phi)\,\chi_{xxy}^{(2)} \right|^2 \\
&+ \left| \cos^2(\phi)\,\chi_{yxx}^{(2)} + \sin^2(\phi)\,\chi_{yyy}^{(2)} + 2\cos(\phi)\sin(\phi)\,\chi_{yxy}^{(2)} \right|^2, \\
=\ &A\cos^4(\phi) + B\cos^3(\phi)\sin(\phi) + D\cos^2(\phi)\sin^2(\phi) + F\cos(\phi)\sin^3(\phi) + G\sin^4(\phi),
\end{split}
\end{equation*}
where $A=|\chi^{(2)}_{xxx}|^2+|\chi^{(2)}_{yxx}|^2$, $B=4(Re(\chi^{(2)}_{xxx}(\chi^{(2)}_{xxy})^*)+Re(\chi^{(2)}_{yxx}(\chi^{(2)}_{yxy})^*))$, $D=2(|\chi^{(2)}_{xxy}|^2+|\chi^{(2)}_{yxy}|^2+Re(\chi^{(2)}_{xxx}(\chi^{(2)}_{xyy})^*)+Re(\chi^{(2)}_{yxx}(\chi^{(2)}_{yyy})^*))$, $F=4(Re(\chi^{(2)}_{xyy}(\chi^{(2)}_{xxy})^*)+Re(\chi^{(2)}_{yyy}(\chi^{(2)}_{yxy})^*))$ and $G=|\chi^{(2)}_{xyy}|^2+|\chi^{(2)}_{yyy}|^2$. The fit in Figure 2b of the main text uses $\phi_0=0.148$ radians, $I_{sub}=0.3$, $A=0.428 \pm 0.005$, $B=0.044 \pm  0.016$, $D=1.296 \pm  0.022$, $F=-0.554 \pm 0.016$, and $G=0.302 \pm 0.006$ ($I_{sub}$, $A$, $B$, $D$, $F$, and $G$ are in arb.~u.). 

The magnetic order is lost above $T_n \sim 45~K$ and both 1L+1L and 2L structures become paramagnetic. $\mathcal{T}$ is restored on average in this phase, which sets $\chi^{odd}_{ijk}=0$. The SHG signal is determined only by the crystal structure and its unitary symmetry operations. The 1L+1L structure belongs to non magnetic PG $m=\mathcal{G}_3=\{\mathcal{I},m_y \}$, and non-zero $\chi_{ijk}^{(2)}$ are determined only by Equation~\ref{eq:R}. The $\chi_{ijk}^{(2)}$ tensor for $\mathcal{G}_3$ in Voigt notation is:
\begin{equation*}
\chi^{(2)}_{\mathcal{G}_3} = 
\begin{pmatrix}
\chi^{(2)\text{ even}}_{xxx} &\chi^{(2)\text{ even}}_{xyy} & \chi^{(2)\text{ even}}_{xzz} & 0 & \chi^{(2)\text{ even}}_{xxz} & 0 \\
0 & 0 & 0 & \chi^{(2)\text{ even}}_{yyz} & 0 & \chi^{(2)\text{ even}}_{yxy} \\
\chi^{(2)\text{ even}}_{zxx} & \chi^{(2)\text{ even}}_{zyy} & \chi^{(2)\text{ even}}_{zzz} & 0 & \chi^{(2)\text{ even}}_{zxz} & 0 \\
\end{pmatrix}.
\end{equation*}

The SHG intensity for $\mathcal{G}_3$ is written as:
\begin{equation*}
\begin{split}
I_{\mathcal{G}_3} (\phi) \propto\ 
&\left| \cos^2(\phi)\,\chi_{xxx}^{(2)\text{ even}} + \sin^2(\phi)\,\chi_{xyy}^{(2)\text{ even}} \right|^2 
+ \left| 2\cos(\phi)\sin(\phi)\,\chi_{yxy}^{(2)\text{ even}} \right|^2, \\
=\ &A\cos^4(\phi) + B\cos^2(\phi)\sin^2(\phi) + D\sin^4(\phi).
\end{split}
\end{equation*}
where $A=|\chi^{(2)\text{ even}}_{xxx}|^2$, $B=2(Re(\chi^{(2)\text{ even}}_{xxx}(\chi^{(2)\text{ even}}_{xyy})^*)+|\chi^{(2)\text{ even}}_{yxy}|^2)$ and $D=|\chi^{(2)\text{ even}}_{xyy}|^2$. We fit the 2L+2L curve shown in Figure 3c above $T_c$  with $I_{sub}=0.2$, $\phi_0=-0.41$ radians, $A=0.018 \pm  0.012$, $B=0.903 \pm 0.049$, and $D=0.778\pm0.012$. ($I_{sub}$, $A$, $B$, and $D$ are in arb.~u.).

\section*{2. Estimation of the out-of-plane electric dipole in 2L+2L \ce{H-CrI3} from experimental result}
We estimate the out-of-plane electric dipole $P_z$ in 2L+2L \ce{H-CrI3} by analyzing the shift of the resistance peaks of the graphene probe, a common approach widely used in other sliding ferroelectric materials$^{26}$, at room temperature (Figure 2e of the main text). Figure \ref{fig:s_dipole}a schematically shows the electrostatic potential profile (cyan curve) across the device when the back gate voltage ($V_b$) is set to 0~V, and thus at the same potential as the top graphene. An electric field arises in the top BN, as there is a potential gradient to compensate for the built-in interlayer potential across the 2L+2L \ce{H-CrI3} ($\Delta V_P$). The electric field in the top BN shifts the charge neutral point of the graphene, which will shift the resistance peak. Figure \ref{fig:s_dipole}b shows the electrostatic potential profile when the back gate voltage is equal to the build-in interlayer potential of the 2L+2L \ce{H-CrI3}  ($V_b=\Delta V_P$). In this case, the built-in interlayer potential exactly compensates for the back gate voltage, meaning there is no voltage drop across the BN layers. This brings the graphene to the charge neutral point, as there is no electric field in the top BN. Therefore, by extracting the locations of the resistance peaks in the forward and backward scans, we can estimate the built-in interlayer potential. As the peaks are not symmetric about $V_b=0$~V, we can determine the built-in interlayer potential by finding the difference between the peak voltages ($\Delta V_b$) and dividing by two: $\Delta V_P$ = $\Delta V_b/2$. 

If the top panel of Figure \ref{fig:s_dipole}c we show the graphene resistance as a function of the bottom gate voltage at 295~K; the same as Figure 2e in the main text, but without converting the voltage to vertical electric field. The peaks in the forward and backward scans are broad, but clearly at different $V_b$. To obtain a better estimate of the peak location, we plot the first derivative of the resistance with respect to the back gate voltage in the bottom panel of Figure \ref{fig:s_dipole}c. The dashed vertical lines indicate where the derivative curves cross the zero point with a negative slope, which correspond to the resistance peaks. As seen in the top panel, the identified peak locations of $V_b=0.38$~V and $V_b=0.78$~V match well with the peaks in the raw resistance data. Having identified the peaks, we find the interlayer potential to be $\Delta V_P = 205$~meV. To estimate the electric dipole $P_z$, we model the polarization as positive and negative charge densities $\pm\rho=\pm P_{2D}/d_c$ separated by the thickness of the 2L+2L \ce{H-CrI3}, $d_c$.
Applying Gauss's law leads to $\rho=\epsilon_cE_c$, where $\epsilon_c$ is the dielectric constant of \ce{CrI3} and $E_c$ is the electric field across the 2L+2L \ce{H-CrI3}. Given the electric field is related to the interlayer potential by $E_c=\Delta V_P/d_c$, we arrive at $P_{z} = \epsilon_c \Delta V_P$. The out-of-plane dielectric constant of \ce{CrI3} is proposed to be $\sim2.7\epsilon_0$$^{41}$. From the extracted built-in interlayer potential, we find $P_{z} \sim 5$~pC/m. Theoretical predictions offer a range of possible values of the polarization of 1L+1L \ce{H-CrI3}. Our experimental measurement of 2L+2L is consistent with larger estimates (i.e. $\sim1.5$~pC/m$^{35}$), which may be due to the thicker system we study (2L+2L) and the broadening of the resistance peaks. Measurements of 1L+1L \ce{H-CrI3} could help clarify the polarization, as well as with monolayer graphene probe and BN encapsulation on both sides to reduce broadening of the peaks$^{26}$.

\section*{3. Ising model simulation for emergent magnetism in \ce{CrI3}}
The intralayer exchange ($J_{\text{intra}}\sim 1.44$ meV) is an order of magnitude larger than the interlayer exchange ($J_{\text{inter}}\sim 0.1$ meV) in few-layer \ce{CrI3}$^{31,32}$, so the interlayer spin alignment is more susceptible to perturbation by $B$-fields. An Ising model with a single spin per layer was adopted here:
\begin{eqnarray}
\label{eq:ising}
    H=-\sum_{i=1}^{N-1}J_{i}S_iS_{i+1}-B\sum_{i=1}^NS_i,
\end{eqnarray}
where a negative sign was placed before the exchange term. $S_i$ represents the magnetization of the $i$-th layer, which can take values $S_i=\pm M_0$, where $M_0=3\mu_B$ is the monolayer magnetization and $\mu_B=5.79 \times 10^{-5}$ eV/T is the Bohr magneton. $J_{i}$ is the exchange coupling between layer $i$ and layer $i+1$. As discussed in the main text, monoclinic stacking interfaces favor AFM interlayer coupling, whereas polar hexagonal stacking interfaces favor FM interlayer coupling. We carry out Monte Carlo (MC) simulations at zero temperature to simulate the hysteresis patterns of the magnetization as the out-of-plane $B$-field is swept back and forth. Each MC step consists of flipping the magnetic moment at a single layer; the change is accepted if the energy decreases compared to the previous state. The Ising magnetization is taken as $\sum_{i}S_i$. 
We use an AFM exchange coupling of $J_{\text{AFM}}M_0^2=-135$ $\mu$eV for natural monoclinic stacking interfaces, which well reproduces the critical $B$-fields of the natural 4L and 6L, as seen in Figure 3a,d in the main text, where the insets show the Ising model simulation. 
To reproduce the experimental curve shown in Figure 3b,e in the main text, we set an FM exchange coupling of $J_{\text{FM}}M_0^2=54$ $\mu$eV for polar hexagonal stacking interfaces. As seen in the insets of Figure 3b,e, the Ising model simulation reproduces the experimental behavior.

\section*{4. Critical field analysis and comparison between natural monoclinic and hexagonal-stacked \ce{CrI3}}
To calculate the critical magnetic field $B_c$ for the first spin-flip transition in 6L CrI$_3$, we use the Ising model (Equation~\ref{eq:ising}) for $N=6$ with AFM couplings $J_i=J<0$ only. The model indicates that the first magnetic transition occurs from $\downarrow\uparrow\downarrow\uparrow\downarrow\uparrow$ (with energy $\varepsilon_{\downarrow\uparrow\downarrow\uparrow\downarrow\uparrow}$) to $\uparrow\uparrow\downarrow\uparrow\downarrow\uparrow$ (with energy $\varepsilon_{\uparrow\uparrow\downarrow\uparrow\downarrow\uparrow}$) in the 6L stack. The transition occurs when the energies of both states are equal. A phenomenological term is included to account for the demagnetization energy, which arises from the internal magnetic field associated with the total magnetization of the sample$^{47,48}$. This contribution is proportional to $M_{\text{bi}}^2/2t$ in CrI$_3$ bilayers, where $M_{\text{bi}}$ denotes bilayer magnetization and $t$ the bilayer thickness. We express the demagnetization energy as $H_{demag}=N_bM^2$, where $N_b$ is the demagnetization constant, and $M$ is the total magnetization of the spin configuration. The energies of both configurations are given by:
\begin{eqnarray*}
    \varepsilon_{\downarrow \uparrow \downarrow \uparrow \downarrow \uparrow} &=& 5JM_0^2, \text{ and}\\
    \varepsilon_{\uparrow \uparrow \downarrow \uparrow \downarrow \uparrow}&=& 3JM_0^2-2M_0B+4M_0^2N_b.
\end{eqnarray*}
The critical field $B_c^{(1)}$ for the first spin-flip transition is:

\begin{eqnarray*}
B_c^{(1)}&=&\frac{-2J_{\text{inter}}+M_{\text{bi}}^2N_b}{M_{\text{bi}}},
\end{eqnarray*}
when $M_{\text{bi}}=2M_0$ and $J_{\text{inter}}=JM_0^2$. 
For the 3L+3L stack Ising model, we use AFM couplings for the natural stacking interfaces ($J_{1,2,4,5}=J<0$) and FM coupling for the polar hexagonal stacking interface ($J_3>0$). 
The first spin-flip transition is from $\downarrow \uparrow\downarrow \downarrow \uparrow \downarrow$ to $\uparrow \uparrow\downarrow \downarrow \uparrow \uparrow$ with total magnetization -2$M_0$ and 2M$_0$, respectively. The energy of both configurations is:
\begin{eqnarray*}
    \varepsilon_{\downarrow \uparrow\downarrow \downarrow \uparrow \downarrow} &=& 4JM_0^2-J_3M_0^2+4M_0^2N_b+2 BM_0,\text{ and}\\
    \varepsilon_{\uparrow \uparrow\downarrow \downarrow \uparrow \uparrow}&=& -J_3M_0^2+4M_0^2N_b-2BM_0.
\end{eqnarray*}
The critical field $B_c^{(2)}$ for this spin transition corresponds to:
\begin{eqnarray*}
\mu_0B_c^{(2)}&=&-\frac{2J_{\text{inter}}}{M_{\text{bi}}}.\\
\end{eqnarray*}
We find a demagnetization term in 6L stacks but not in 3L+3L ones. Since $N_b>0$, this term increases the critical field and may explain the smaller $B_c$ in 3L+3L stacks shown in Figures 3d,e of the main text.

\section*{5. Spin-polarized charge transfer model in 2L+2L \ce{H-CrI3}  under electric fields}
We now examine the magnetization hysteresis in a 2L+2L stack under $E$-sweeping with $B=0$ T. 
The ground state obtained from our Ising model  (Equation~\ref{eq:ising} for $N=4$, with $J_1,J_3<0$ and $J_2>0$) suggests a magnetic ground state  $\uparrow \downarrow\downarrow \uparrow$. However, there may still be a small finite magnetization in absence of $E$, due to asymmetric charge distribution between the layers, the effect of the substrate creating an inequivalence between the layers, and the out-of-plane electric dipole ($P_z$) redistributing charge. We model this effect as:
\begin{eqnarray*}
    S_1(P_z)&=&S_1(1+(1-a)\eta P_z),\\
    S_2(P_z)&=&S_2(1+a\eta P_z),\\
    S_3(P_z)&=&S_3(1-b\eta P_z),\\
    S_4(P_z)&=&S_4(1-(1-b)\eta P_z),
\end{eqnarray*}
where $\eta$ is the coupling constant and $a,b$ represent the asymmetric charge redistribution between the top and bottom layers. The ground state has no net magnetization in the symmetric case ($a=b$). A net magnetization of $2(a-b)M_0\eta P_z$ arises when $a>b$, whereas the opposite magnetic state ($\downarrow\uparrow\uparrow\downarrow$) exhibits a magnetization of $-2(a-b)M_0\eta P_z$. 

The charge is redistributed within the layers when $E$ is applied upward. We consider a 4L CrI$_3$ stack encapsulated by hBN with graphene gates to model this redistribution. The hBN dielectrics are symmetric in the model, each having a width $d$ and dielectric constant $\varepsilon$. The interlayer distances are $t$ and $t_2$ for the natural and polar stacks, with dielectric constants $\varepsilon_1$ and $\varepsilon_2$. Vertical electric fields in the hBN, natural and polar stacks are denoted $E_0,E_1$ and $E_2$, respectively. The dielectric environment differs between stacking orders, giving different $\varepsilon$ values for natural and polar stack. Details of the setup can be seen in Figure \ref{fig:capacitor}. The induced charge in the layers is obtained from Gauss’s law:
\begin{eqnarray*}
    n_1e&=&\varepsilon_0(\varepsilon E_0-\varepsilon_1E_1), \\
    n_2e&=&\varepsilon_0(\varepsilon_1 E_1-\varepsilon_2E_2), \\
    n_3e&=&\varepsilon_0(\varepsilon_2 E_2-\varepsilon_1E_1), \\
    n_4e&=&\varepsilon_0(\varepsilon_1 E_1-\varepsilon E_0),
\end{eqnarray*}
where $n_i$ is the charge density induced by $E$ in layer $i\in [1,...,4]$, $e$ the elementary charge and $\varepsilon_0$ the vacuum dielectric constant. We note $n_1=-n_4$, and $n_3=-n_4$. We use this result to write the spin transformation under $E$ as:
\begin{eqnarray*}
    S_1(E)&=&S_1(1+\alpha_1E+(1-a)\eta P_z(E)),\\
    S_2(E)&=&S_3(1+\alpha_2E+a\eta P_z(E)),\\
    S_3(E)&=&S_4(1-\alpha_2E-b\eta P_z(E)),\\
    S_4(E)&=&S_5(1-\alpha_1E-(1-b)\eta P_z(E)),
\end{eqnarray*}
where $\alpha_1,\alpha_2$ are coupling constants. The magnetization of the $\uparrow\downarrow\downarrow\uparrow$ stack under $E$ is:
\begin{equation}\label{eq:mag}
 M(E)=2(a-b)\eta M_0 P_z(E).
\end{equation}
Note that Figure 4d in the main text presents the raw RMCD change and fractional change of the magnetization relative to the RMCD value of one of the bistable states (state $A$), 
to highlight the non-volatile $\sim3\times$ magnetization change between the two states. 
To compare the magnetization change with the Ising model, in Figure \ref{fig:LFE-model}a we plot the experimental magnetization as $\Delta M_{\text{exp}} = (M-M_A)/|M_{\text{sat}}|$, where $M$ is the RMCD, $M_A$ is the RMCD of state $A$, and $|M_{\text{sat}}|=18.5\%$~RMCD is the saturation RMCD value of the FM state (see Fig. 3b of the main text), as a function of $E$. Normalizing with respect to the total magnetization enables a direct comparison between experiment and theory. 

We model $P_z(E)$ using the Landau free energy function given by:
\begin{equation}\label{eq:free_energy}
F(P_z)=\alpha P_z^2+\beta P_z^4-\gamma EP_z,
\end{equation}
where $\alpha=$, $\beta$, and $\gamma$ are constants to be determined from \textit{ab initio} calculations or experiments. $P_z$ is obtained by solving $\partial F(P_z)/\partial P_z=0$. If $\pm P_{0z}$ denotes the two local minima of $F(P_z)$, Equation \ref{eq:free_energy} is written as:
\begin{equation*}
    F(P_z)=\beta(P_z^4-2P_{0z}^2 P_z^2)-\gamma EP_z.
\end{equation*}
We use parameters obtained from DFT calculations$^{34}$ of a 1L+1L stack: $P_{0z}=0.3$ pC/m, $\beta=5.5$ eV m$^4$/pC$^4$ and $\gamma=3.43$ nm$^2$. $P_z$ was obtained by solving $\partial F(P_z)/\partial P_z=0$. This provides a qualitative understanding of the 2L+2L case, which may have a larger dipole and a different sliding barrier. Figure \ref{fig:LFE-model}b shows $F(P_z)$ without $E$. The system remains trapped in one minimum until it loses stability at the coercive field $E_c$, which triggers a switch in polarization. When the field is swept in the opposite direction, the same behavior occurs at $-E_c$. These features reflect the presence of an energy barrier between the two minima. We numerically minimize $F(P_z)$ using a home-made Python code to obtain $P_z(E)$. The minimization is initialized with a seed close to the expected solution, ensuring convergence to a local minimum. The solution obtained for a given $E$-field is then used as the initial seed for the next field value, allowing us to follow the metastable branches as the field is ramped up or down. This procedure naturally captures the ferroelectric barrier and can reproduce the ferroelectric hysteresis loop. Figure \ref{fig:LFE-model}c shows $P_z(E)$ obtained from this process, which captures the hysteretic behavior. 

To model magnetization in the 2L+2L stack under $E$, we include the $P_z(E)$ curve obtained from the free-energy model and the minimization process into Equation \ref{eq:mag}. Figure \ref{fig:LFE-model}d shows the resulting magnetization curve, which displays a hysteresis driven by sliding ferroelectricity. To obtain a magnetization change of the same order as the experimental value, we set $2(a-b)\eta = 0.14$ m/pC. This approach suggests a $P_z$ flipping (or ferroelectric transition) that may explain the hysteresis seen in Figure 4f in the main text.

\newpage
\begin{figure}
    \centering
    \includegraphics[width=\linewidth]{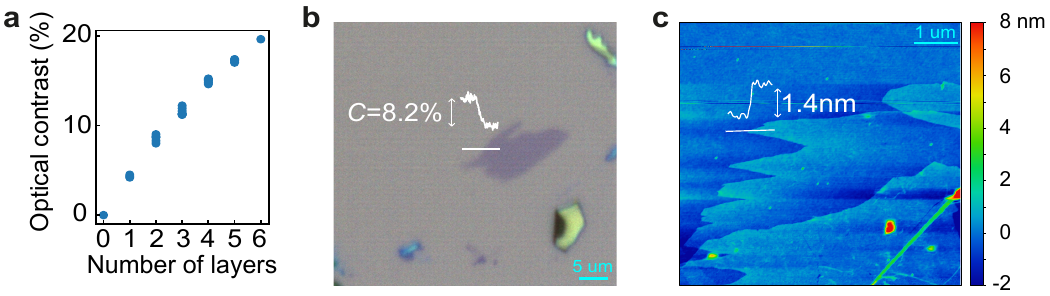}
    \caption{\textbf{Identifying the layer number of \ce{CrI3} flakes.} \textbf{a,} Optical contrast {\em vs.}~number of layers, based on several flakes of different thicknesses. The contrast is defined as $C = \frac{I_s-I_f}{I_s+I_f}$, where $I_s$ and $I_f$ are the intensities of the substrate and flake, respectively. \textbf{b,} Optical image of a flake, with the line cut showing a $8.2\%$ contrast, consistent with a bilayer. \textbf{c,} Atomic force microscopy image of a stack made from the flake in \textbf{b}. The inset shows a height of 1.4 nm, consistent with that of a bilayer (a layer has a thickness of 0.7 nm).}
    \label{fig:s1}
\end{figure}

\begin{figure}
    \centering
    \includegraphics[width=\linewidth]{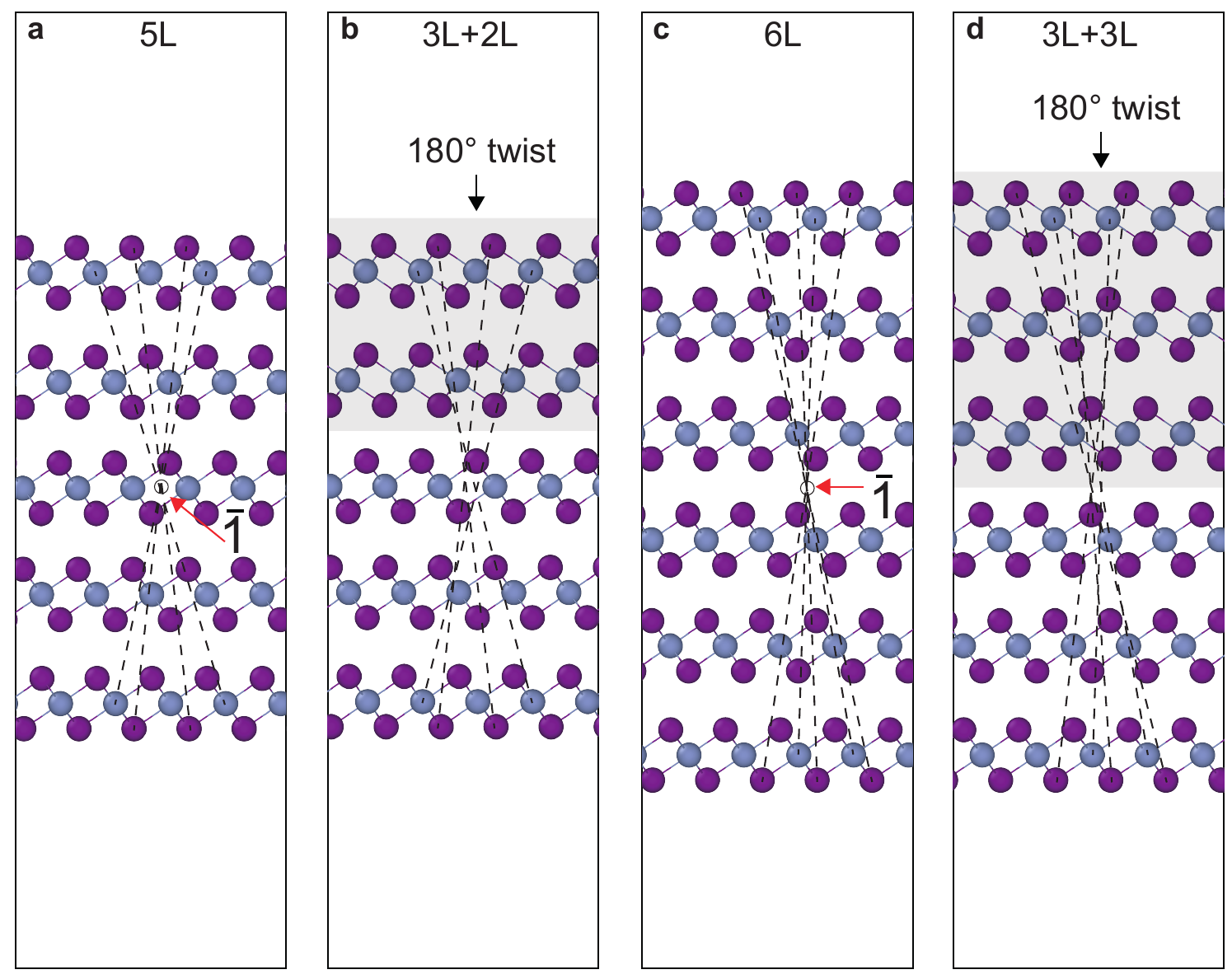}
    \caption{\textbf{Additional few-layer \ce{CrI3} crystal structures.} \textbf{a-b}, Side views of (a) 5L and (b) 3L+2L \ce{CrI3} crystals. \textbf{c,d}, Side views of 6L and 3L+3L \ce{CrI3} crystals. Grey shading marks the 180$^\circ$ twisted layers. The center of inversion in paramagnetic phases is represented by a black circle in 5L and 6L stacks. The purple balls represent \ce{I} atoms and the blue balls represent \ce{Cr} atoms.}
    \label{fig:s3}
\end{figure}

\begin{figure}
    \centering
    \includegraphics[width=\linewidth]{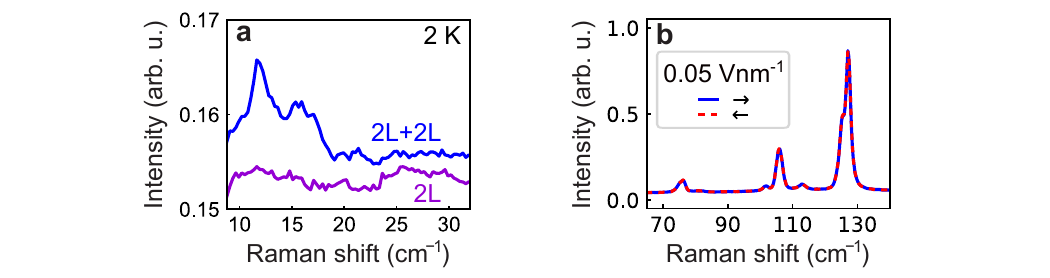}
    \caption{\textbf{Additional Raman spectrum at low temperature and with finite electric field at 2~K.} \textbf{a-b}, Ultralow-frequency Raman spectrum at 2~K. Low-frequency modes in the 2L+2L \ce{H-CrI3} (blue curve) persist at low temperature, while the natural 2L stack (purple curve) still shows no relevant peaks. \textbf{b}, Raman spectrum of the 2L+2L \ce{H-CrI3} used in Figure 4d-e of the main text at $V_b=0.05$V~nm$^{-1}$ (where there is large hysteresis in Figure 4d-e of the main text) during the ascending (blue) and descending (dashed red) part of a back gate sweep. The identical spectra suggests intralayer structural changes are not responsible for the gate-dependent behavior attributed to polarization switching in the main text.}
    \label{fig:raman-temp}
\end{figure}

\begin{figure}
    \centering
    \includegraphics[width=\linewidth]{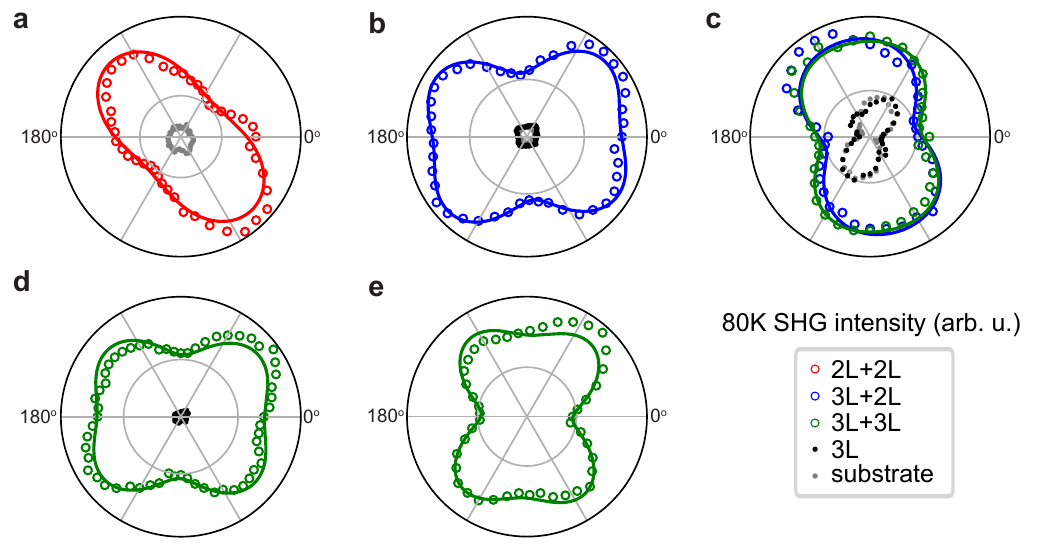}
    \caption{\textbf{SHG measurements of additional {\ce{H-CrI3}} samples at 80~K.} \textbf{a-e}, SHG as a function of incident linear polarization angle for a 2L+2L stack (a), two 3L+3L stacks (b,c), and two 3L+2L stacks (d,e). Solid lines are theoretical fits to the second order non-linear susceptibility tensor of the $m$ PG. The data were taken at 80~K, well above $T_N$, to highlight the polar stacking contribution and remove any magnetic order contribution. The legend in the bottom right is for all panels.}
    \label{fig:s4}
\end{figure}

\begin{figure}
    \centering
    \includegraphics[width=\linewidth]{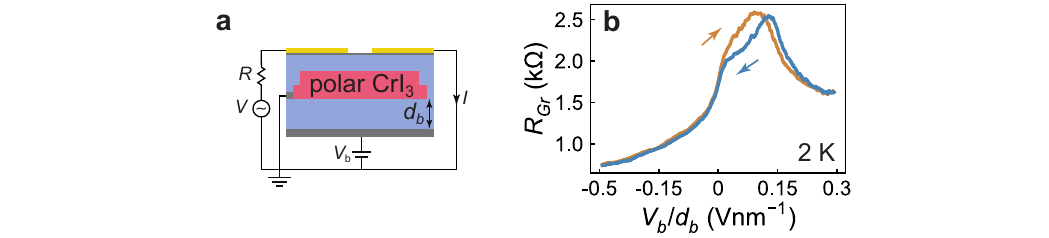}
    \caption{\textbf{Ferroelectric behavior in another \ce{H-CrI3} device.} \textbf{a,} Schematic of a device where few-layer graphene probes the local $E$ in the top dielectric layer, which is sensitive to $\mathbf{P}$-switching of the polar stacked \ce{CrI3}, and a graphite contact is used to ground the \ce{CrI3}. \textbf{b,} Top graphene resistance ($R_{Gr}$) as the bottom gate is swept up and down in a loop for a 2L+2L stacked device at 2 K, showing hysteretic behavior indicative of sliding ferroelectricity. This device was not large enough for optical measurement.}
    \label{fig:s_rgr_d4}
\end{figure} 

\begin{figure}
    \centering
    \includegraphics[width=\linewidth]{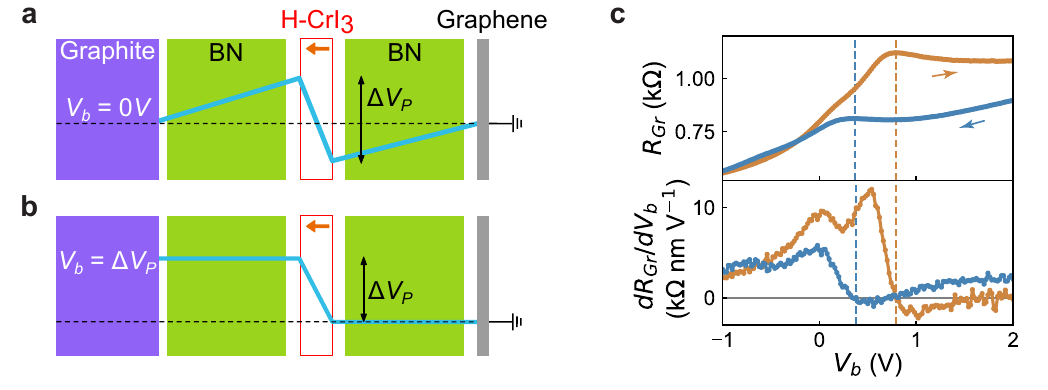}
    \caption{\textbf{Estimating the out-of-plane electric dipole in 2L+2L \ce{H-CrI3}.} \textbf{a,} Schematic illustration of the electrostatic potential profile (cyan curve) across the device shown in Figure 2d of the main text when $V_b=0$~V. \textbf{b,} The potential energy profile when $V_b$ is equal to the built-in interlayer potential of the 2L+2L \ce{H-CrI3} ($\Delta V_P$), where the top graphene is at the charge neutral point. \textbf{c,} The top panel shows the resistance of the graphene ($R_{Gr}$) as $V_b$ is swept in a loop at 295~K. The bottom panel shows the first derivative of $R_{Gr}$ with respect to $V_b$. Dashed vertical lines indicate the resistance peak locations identified for the forward and backward sweeps.}
    \label{fig:s_dipole}
\end{figure} 

\begin{figure}
    \centering
    \includegraphics[width=0.5\linewidth]{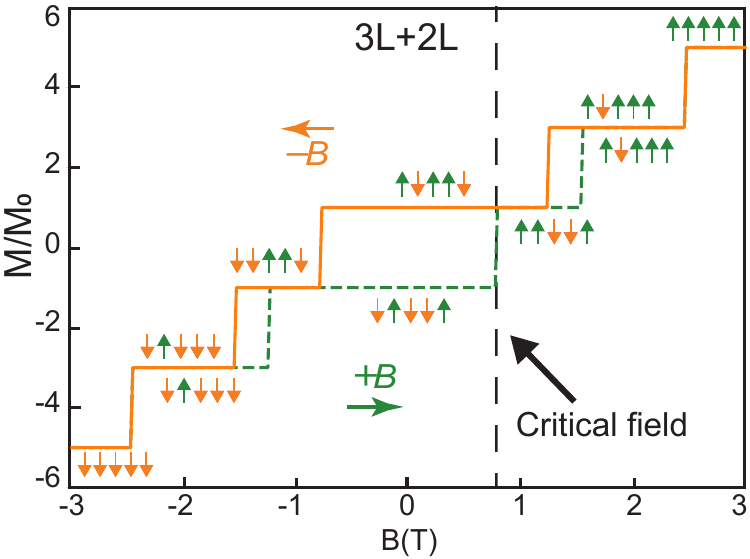}
    \caption{\textbf{Magnetic hysteresis modeling in 3L+2L \ce{H-CrI3}.} Magnetization obtained from a five-spin Ising model (Equation \ref{eq:ising}) with AFM couplings $J_{1,2,4}=-135$ $\mu$eV for a natural interface and with FM coupling $J_3=54$ $\mu$eV at the polar hexagonal interface for a 3L+2L stack. A demagnetization term is included as $H_{\text{demag}}=N_b\left(\sum_i S_i\right)^2$ with $N_b=20$ $\mu$eV/$\mu_B^2$. The first spin-flip transition marked in the plot corresponds to the $E$-field switching shown in Figure 4c of the main text.}
    \label{fig:s_hysteresis}
\end{figure}

\begin{figure}
    \centering
    \includegraphics[width=0.5\linewidth]{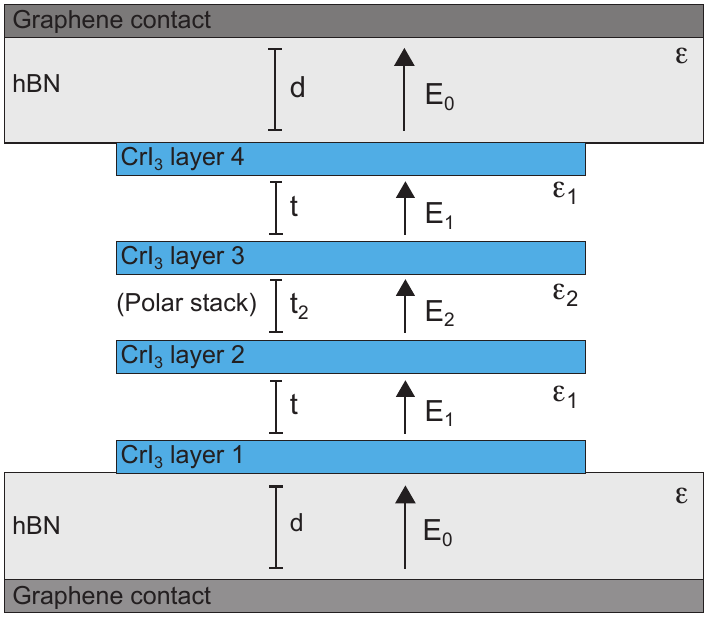}
    \caption{\textbf{capacitor model for 2L+2L \ce{H-CrI3}.} Schematics of a 2L+2L stack encapsulated by hBN, with graphene top and bottom gates.}
    \label{fig:capacitor}
\end{figure}

\begin{figure}
    \centering
    \includegraphics[width=\linewidth]{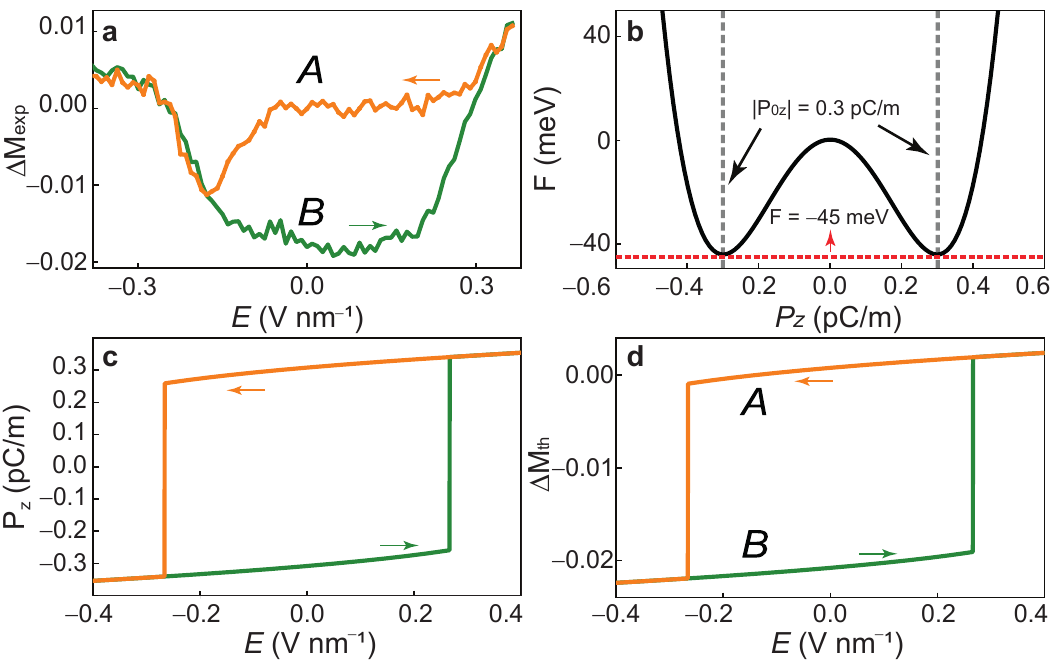}
    \caption{\textbf{Sliding ferroelectric model simulation for 2L+2L \ce{H-CrI3}.} \textbf{a}, Out-of-plane electric field control of magnetism at $B=0$ T for a 2L+2L \ce{H-CrI3}. $\Delta M_{\text{exp}}=(M-M_A)/|M_{\text{sat}}|$, where $M$ is the measured RMCD, $M_A$ is the RMCD of state $A$, and $|M_{\text{sat}}|=18.5\%$ is the saturation RMCD in the FM state. \textbf{b}, Free-energy model, where red horizontal dashed line represents the energy barrier and vertical dashed line are the two local minima at $P_z = 0.3$ pC/m. We parametrize this curve from DFT calculations of a 1L+1L stack, which provides a qualitatively understanding of the 2L+2L case, which may have larger dipole and different sliding barrier.  \textbf{c}, Polarization hysteresis obtained from the free-energy model. \textbf{d}, Theoretical zero-field ($B=0$ T) magnetization $\Delta M_{th}=(M(E)-M_A)/4M_0$ of the ground state $\uparrow \downarrow\downarrow\uparrow$ normalized by the total magnetization ($4M_0$) in FM state. $M_A$ correspond to the magnetization of the state A and the polarization contribution is introduced from the results of panel c.}
    \label{fig:LFE-model}
\end{figure}

\end{document}